\documentclass[12pt]{article}
\pdfoutput=1
\usepackage{euscript,amsmath,amsthm,amsfonts,amssymb}
\usepackage{mathtools}
\usepackage{subfig}
\usepackage{float}
\usepackage[margin=1in]{geometry}
\usepackage{graphicx}
\usepackage{outlines}
\usepackage[dvipsnames]{xcolor}
\usepackage{jheppub}

\DeclareMathOperator{\area}{area}

\title{\boldmath Multipartite entanglement and topology in holography}
\author{Jonathan Harper}
\affiliation{Martin Fisher School of Physics, Brandeis University, Waltham, Massachusetts 02453, USA}
\preprint{BRX-TH-6664}
\emailAdd{jharper@brandeis.edu}
\abstract{Starting from the entanglement wedge of a multipartite mixed state we describe a purification procedure which involves the gluing of several copies. The resulting geometry has non-trivial topology and a single oriented boundary for each original boundary region. In the purified geometry the original multipartite entanglement wedge cross section is mapped to a minimal surface of a particular non-trivial homology class. In contrast, each original bipartite entanglement wedge cross section is mapped to the minimal wormhole throat around each boundary. Using the bit thread formalism we show how maximal flows for the bipartite and multipartite entanglement wedge cross section can be glued together to form maximal multiflows in the purified geometry. The defining feature differentiating the flows is given by the existence of threads which cross between different copies of the original entanglement wedge. Together these demonstrate a possible connection between multipartite entanglement and the topology of holographic spacetimes.

}
\begin{document}
\maketitle
\flushbottom

\section{Introduction}
The connection between geometry and entanglement has been a fruitful tool of modern research. This is most clearly evidenced by the Ryu-Takayanagi (RT) formula \cite{Ryu_2006} and its covariant generalization \cite{Hubeny_2007} which allows one to relate the area of a bulk minimal surface and the entanglement entropy of the boundary of a holographic CFT. More recently, efforts have been made to generalize this to other interesting surfaces in the bulk, most notably the entanglement wedge cross section which is known to contain information about quantum and classical correlations and acts as a measure of mixed state entanglement \cite{Umemoto_2018,Nguyen2018,Bao2018}. The entanglement wedge can be purified by gluing a copy of the CPT conjugate and gluing along the common boundaries. In the purification the entanglement wedge cross section is related to the entanglement entropy between the two sides of the constructed wormhole.

An important generalization can be performed by considering more than two boundary regions. One then considers the multipartite entanglement wedge cross section whose area should be thought of as a measure of multipartite entanglement in the CFT \cite{Umemoto2018,Bao:2018aa}. The question remains how one should think of purifications in this case. A simple doubling of the geometry does not have the effect of relating the topological data, the area of minimal surfaces in each homology class of the manifold, to that of the multipartite entanglement wedge cross section. In this paper we will do just this: using several copies of the original entanglement wedge we will illustrate a gluing procedure which will result in a particular purification where the multipartite entanglement wedge cross section can be related to a minimal surface in a particular homology class of the manifold. This restores the relationship between topology and entanglement present in the bipartite case.

We will find it useful to use the bit thread formalism \cite{Freedman2017,Headrick_2018} which allows one to relate geometric surfaces to maximal, normed, divergenceless vector fields, or flows, in the bulk. It was shown that both the bipartite \cite{Harper:2019lff, Du:2019emy} and multipartite \cite{Harper:2019lff} entanglement wedge cross section can be written in bit thread language. Using these flows as building blocks we can use them to create a new flow on the purified geometry. Importantly, these flows originally can end on all boundaries of the entanglement wedge. Since these are glued to create the purified manifold we will find these flows can be joined across the boundaries. Using the tools of convex optimization we will describe how to calculate these flows directly using the topology of the manifold.

In section \ref{sec:setup} we give some background detailing the multipartite entanglement wedge cross section and bit thread duals. Next, in section \ref{sec:BTwarmup} we describe how mixed states generally contain bit threads which end on boundaries and show how gluing the geometry to form a pure state allows the threads on each copy to be identified to form a single flow. We also show how convex duality, which relates geometrically surfaces to flows, can be performed using knowledge of the homology class of the surface. Then, in section \ref{sec:mreflected} we put these together to describe a purification procedure for the entanglement wedge of a multipartite mixed state. Using our construction we will find the multipartite entanglement wedge cross section can be described as a minimal surface in a particular homology class of the manifold and thus is related to topological data. Using the tools developed we will describe flows of bit threads which calculate this quantity and are built from the original flows which calculated the multipartite entanglement wedge cross section. Finally, in section \ref{sec:discuss} we describe some connections between our construction and multipartite reflected entropy, tensor networks, and multiboundary wormholes. To conclude we emphasize a possible connection between measures of multipartite entanglement and the topology of holographic spacetimes.

While in preparation the papers \cite{Bao:2019zqc,Chu:2019etd} appeared which describe different approaches for constructing purifications of the multipartite entanglement wedge cross section. Notably, both understood the need for multiple copies of the original geometry in these constructions.
 
\section{Holographic multipartite entanglement wedge cross section}\label{sec:setup}
In this section we review the geometric construction of the multipartite entanglement wedge cross section and corresponding bit thread programs. These surfaces and their dual flows will be used later to motivate our procedure. Throughout this paper we will be primarily working with static slices of pure $AdS_{3}$. However, both the entanglement wedge cross section and bit thread can be used in the context of higher-dimensional and covariant geometries as well as excited states \cite{Umemoto_2018,Freedman2017,Headrick_2018,cbt,cewc}.
\subsection{Geometric construction}

\begin{figure}[H]
\centering
\includegraphics[width=.4\textwidth]{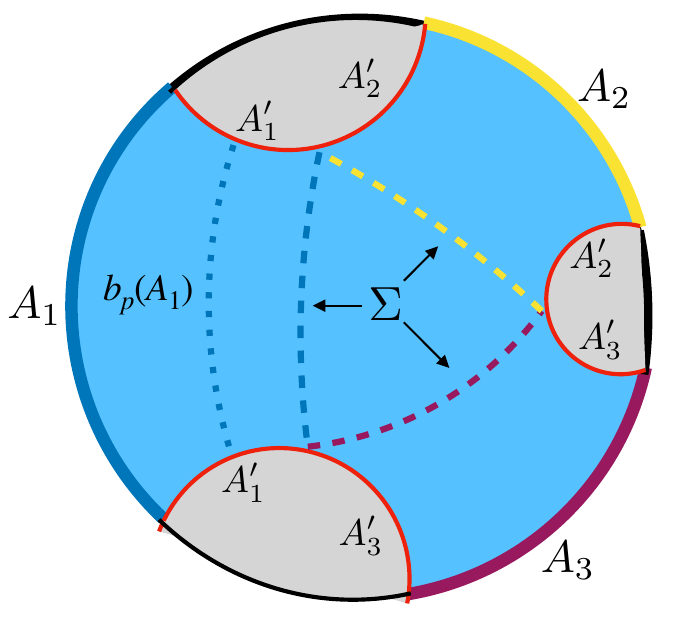}
\caption{\label{fig:multEOP}
The geometric construction of the bipartite and multipartite entanglement wedge cross sections illustrated for the case of three regions. The joint RT surface $\mathcal{O}\coloneqq m(\mathcal{A})$ is shown in red. This is then partitioned into regions $A_i'$, and the total area of the corresponding minimal surfaces $m(A_iA_i')$ is minimized over all partitions. The minimal surfaces $\Sigma(A_{i})$ are shown as colored dashed lines; their union is $\Sigma$. Also shown as a dotted line is the surface $b_p(A_1)$ which is the minimal surface homologous to $A_{1}$ relative to $\mathcal{O}$.
}
\end{figure}

Starting with a static time slice of AdS we chose $n$ non-overlapping sections of the boundary $\mathcal{A} = \cup_{i}^{n} A_{i}$. We then calculate the bulk minimal surface $\mathcal{O}\coloneqq m(\mathcal{A})$ whose area, using the RT formula, would give the entanglement entropy $S(\mathcal{A})$. We can now define the multipartite entanglement wedge, $r(\mathcal{A})$, as the bulk region with boundary $\mathcal{A} \cup \mathcal{O}$. In this region there are two important classes of surfaces we can consider:

 The bipartite entanglement wedge cross section $b_{p}(A_{i})$ can be calculated as the minimal surface homologous to $A_{i}$ relative to $\mathcal{O}$ for each $A_{i}$ \cite{Umemoto_2018,Nguyen2018}. This has the effect of allowing the surfaces to begin and end on $\mathcal{O}$. The union of these minimal surfaces we will call $b_{p}$.

To determine the multipartite entanglement wedge cross section, $\Sigma$, we instead start by partitioning $\mathcal{O}$ into regions  $A_i'$. $\Sigma$ is then given as the union of $m(A_iA_i')$ minimized over all such partitions \cite{Umemoto2018,Bao:2018aa}. This construction is summarized in figure \ref{fig:multEOP}.

\subsection{Bit thread duals}
We will now review how $b_{p}$ and $\Sigma$ can be described in terms of bit threads \cite{Harper:2019lff}. The bit thread formalism is a rewriting of the RT formula making use of Lagrange duality and other tools of convex optimization. The entanglement entropy is calculated as the maximum flux over all divergenceless, normed vector fields. We can think of the integral curves of this vector fields as ``threads" which connect two boundary regions. Such a choice of threads is usually thought of a distillation of bell pairs whose number gives precisely the entanglement entropy  \cite{Freedman2017,Headrick_2018}. In order to consider quantities other than the entanglement entropy one must necessarily modify the bit thread optimization program. These changes can also be derived making use of Lagrange duality applied to the original minimization program in terms of surfaces\footnote{For those wishing for a more thorough review of bit threads see both \cite{Freedman2017,Headrick_2018}.}.

\begin{figure}[H]
\centering
\includegraphics[width=.4\textwidth]{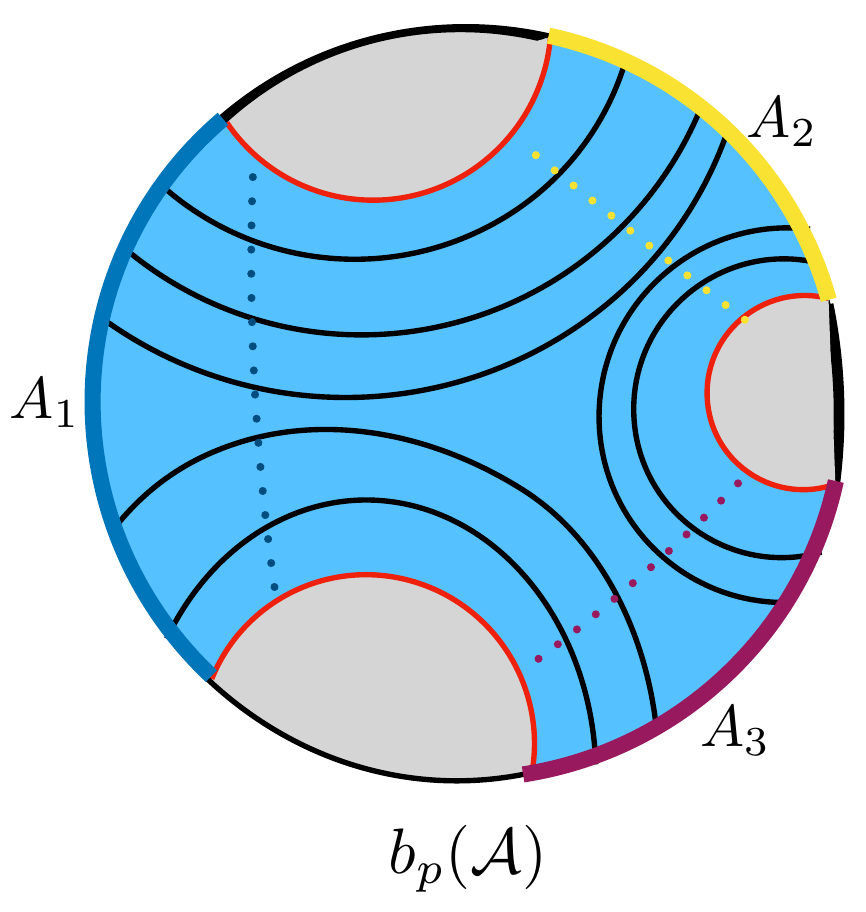}
\caption{\label{fig:bp}The sum of the area of the bipartite homology region cross sections for multiple regions is given by the flux of a maximal multiflow confined to $r(\mathcal{A})$. }
\end{figure}
To calculate $b_{p}$ we consider a maximal flow connecting all $n$ boundary regions (a ``maximal multiflow" \cite{Cui:2018aa}) which is confined to $r(\mathcal{A})$:
\begin{equation}
\begin{split}
\area(b_{p}) \coloneqq \max_{\mathcal{V}} \sum_{i,\; j}\int_{A_{i}} \sqrt{h}n_{\mu}v^{\mu}_{ij} \text{ such that } n_{\mu}v_{ij}^{\mu}|_{\mathcal{O}} =0, \; \nabla_{\mu}v_{ij}^{\mu} = 0\,, \; \sum_{j}|v_{ij}| \leq 1
\end{split}
\end{equation}
where $\sqrt{h}$ and $n_{\mu}$ are respectively the induced metric and normal covector on the boundary. Each $v_{ij}$ is a vector field with flux only on $A_{i}$ and $A_{j}$. Such a flow maximizes the total number of threads connecting all $n$ boundary regions which can be placed in the entanglement wedge. The natural geometric obstruction to such flows in the bulk is exactly $b_{p}$ (see figure \ref{fig:bp}).

The flow program for $\Sigma$ is more involved:
\begin{equation}\label{meopintro}
\begin{split}
\area(\Sigma)= \max_{\mathcal{V},\alpha} \left(\sum_{i}\int_{A_{i}}\sqrt{h}\, n_{\mu}v^{\mu}_{i} + \int_{\mathcal{O}}\sqrt{h}\,\alpha \right) \\ \text{such that}\quad\nabla_{\mu}v_{i}^{\mu} = 0\,, \; |v_{i}| \leq 1\,, \;\left.n_{\mu}v^{\mu}_{i}\right|_{\mathcal{O}} = \alpha\,
\end{split}
\end{equation}

\begin{figure}[H]
\begin{tabular}{ccc}
\includegraphics[width=.3\textwidth]{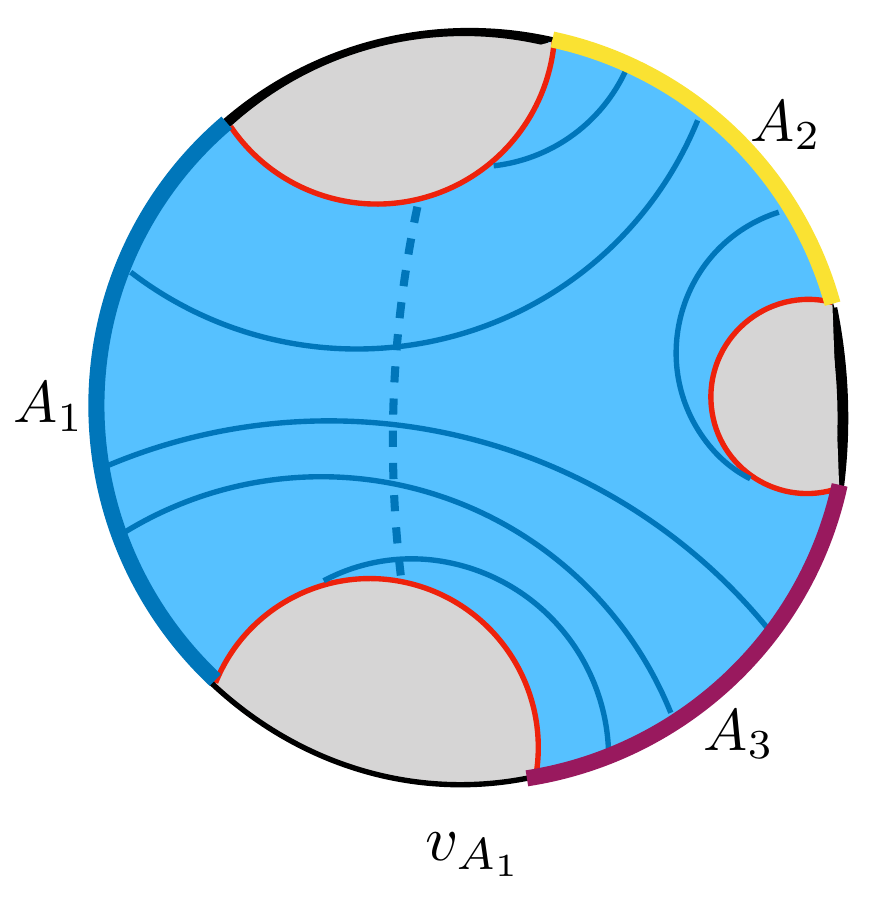} & \includegraphics[width=.3\textwidth]{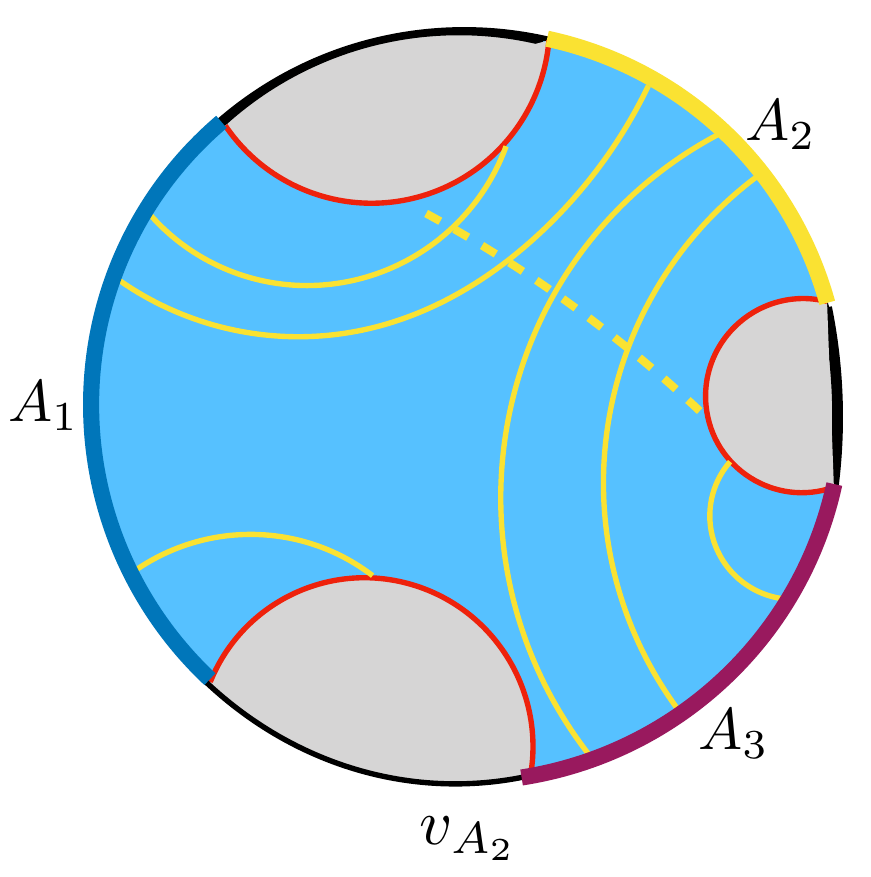} & \includegraphics[width=.3\textwidth]{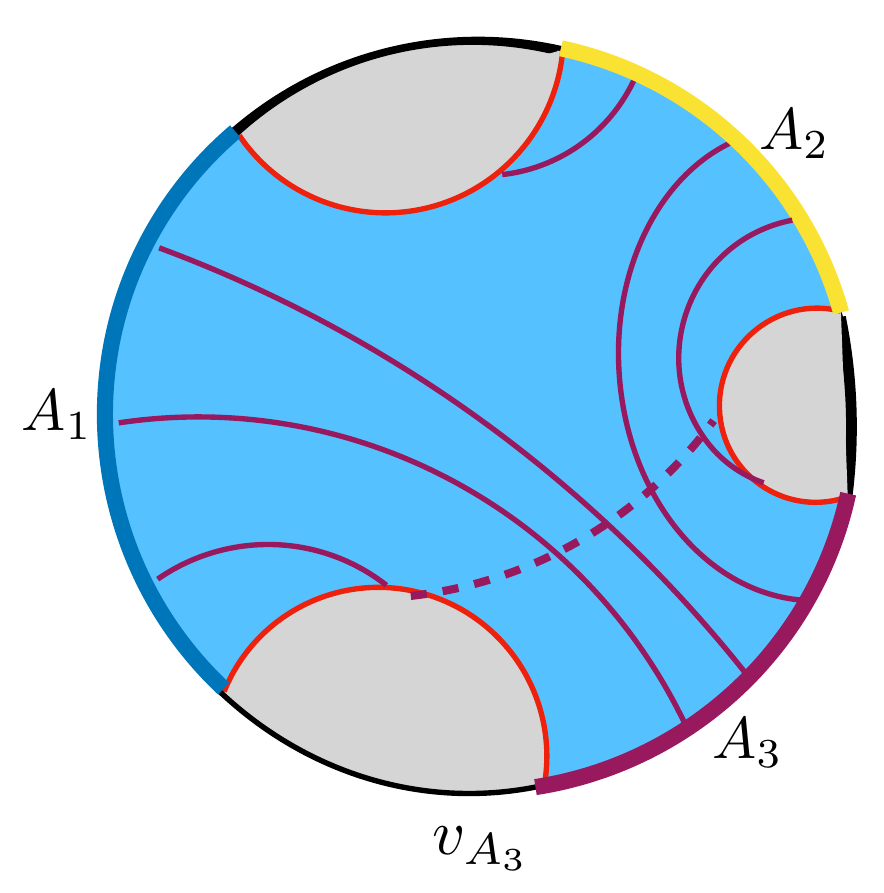}\\
\end{tabular}
\caption{\label{fig:meopexample}An example of a maximal flow configuration $\mathcal{V}$ to \eqref{meopintro} for three regions. Note the local flux on $\mathcal{O}$, which we call $\alpha$, is the same for all three of the flows. In the example each surface $b_{p}(A_{i})$ can support three threads while each surface $\Sigma(A_{i})$ can support four.}
\end{figure}

because it utilizes several flows, one for each boundary region $A_i$\footnote{Unlike the previous case these flows together \emph{do not} form a multiflow. This is because the vector fields do not necessarily satisfy a joint norm bound. Said another way if we to attempt to place all of these flows together on the same manifold the norm bound would generically be violated.}. These flows interact with each other only through the joint constraint $n_\mu v_i^\mu|_{\mathcal{O}}=\alpha$. The flow lines can be thought of as different-colored threads living on individual copies of the manifold. As an example of an optimal configuration see figure \ref{fig:meopexample}.

An important feature of \eqref{meopintro} is that at no point were we required to define a partition of $\mathcal{O}.$ That is the threads are able to find the correct location of $\Sigma$ (where they saturate) naturally as a result of the maximization. This is in contrast to the cut side where we were required to explicitly minimize over both the area of surfaces and choice of partition. 

It is important to note that maximal thread configurations are generically non-unique and it is often the case that among these exist certain configurations with ``nice" properties; this is not unlike a choice of gauge. To implement this choice we can impose additional restrictions on the maximal program \eqref{meopintro} in such a way that these do not change the value of the objective. We choose to require the flux on $\mathcal{O}$ to be positive, but as small as possible without changing the maximum value of the objective. This then guarantees that all of the threads of $v_{i}$ which begin on $A_{i}$ will pass through and saturate on $b_{p}(A_{i})$. The additional threads sourced from $\mathcal{O}$ then provide the extra flux necessary for the flow to saturate on $\Sigma_{A_{i}}$. This allows for the interpretation that the threads from $\mathcal{O}$ which couple the flows $v_{i}$ together are ``truly" multipartite in the sense that because they must be sourced from $\mathcal{O}$ they can not contribute to the saturation of $b_{p}$  \cite{Harper:2019lff}.

\section{Biparite mixed states, purification, and antiloop dualization}\label{sec:BTwarmup}

An important feature of holographic mixed states is the existence of boundaries besides the CFT boundary. These play an important role of changing the homology class with which the minimal surface is calculated. On the flow side duality shows this is equivalent to allowing bit threads to end on these boundaries. Thus generically such states will have bit thread configurations which end there. When performing a purification these boundaries must be eliminated which is usually done by gluing one geometry to another along these surfaces. In the simplest cases this is usually the CPT conjugate of the original geometry. On the purification we can consider two different methods for generating valid bit thread configurations: Constructing a maximal thread configuration on each copy we then purify and identify threads connecting each side of the glued boundary. We call such flows ``constructed flows". Alternatively we can purify and then directly optimize on the new purified manifold giving us an ``optimized flow". In what follows we will find both view points helpful.

\subsection{Purifying bipartite mixed states}

\begin{figure}[H]
\centering
\includegraphics[width=.4\textwidth]{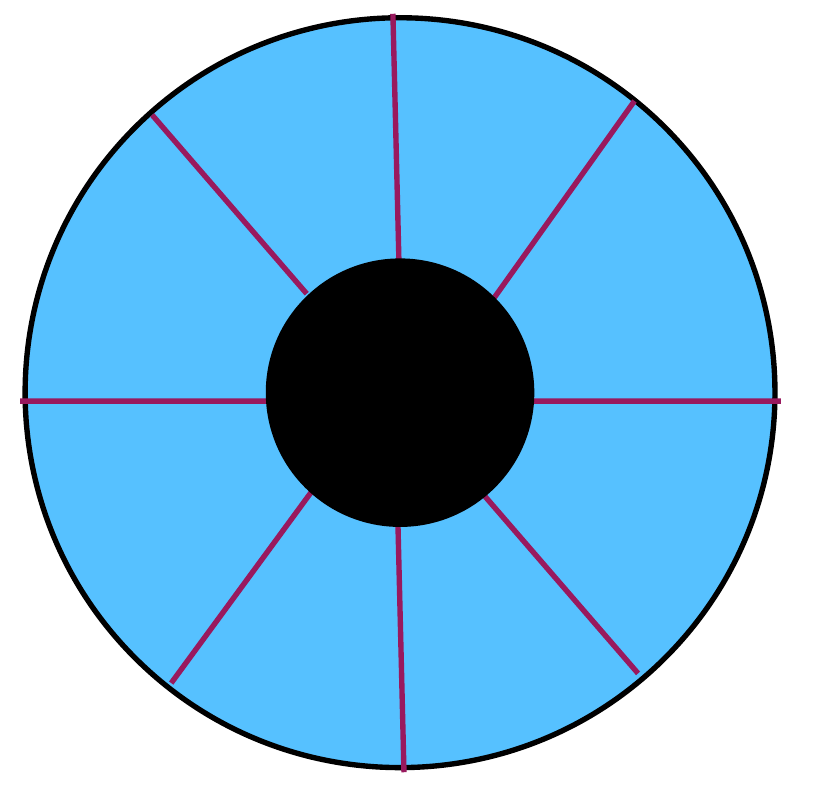}
\caption{\label{fig:bhthreads} A maximal thread configuration connects the CFT boundary to the blackhole horizon where it saturates.}
\end{figure}

\subsubsection{Thermofield double}
As a simple illustrative example we consider a blackhole geometry in AdS$_{3}$ which is dual to a thermal mixed state of the boundary CFT. We take our region to be the entire boundary and calculate the entanglement entropy. It is well known that the minimal surface will be at the blackhole horizon. This is due to the homology constraint when minimizing to find the RT surface. As a result the maximizing bit thread configuration is non zero: the threads start on the CFT boundary and end on the minimal surface, the blackhole horizon (see figure \ref{fig:bhthreads}). We will see that this is a salient feature of mixed state and later multipartite entanglement in the bit thread language, often threads will end on boundaries in the geometry other than the CFT boundary.

\begin{figure}[H]
\centering
\includegraphics[width=.5\textwidth]{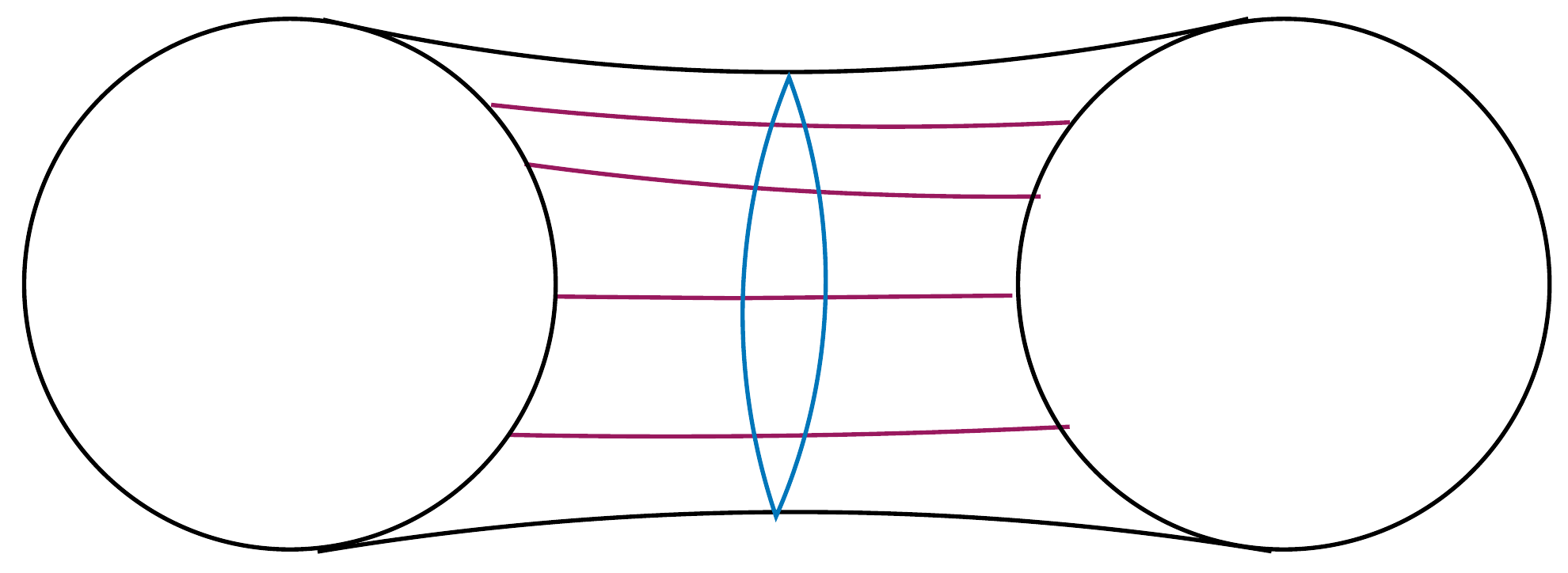}
\caption{\label{fig:tfdthreads} Upon purifying the threads can naturally be seen as extending from one side of the wormhole to the other saturating on the minimal wormhole throat.}
\end{figure}

Next, we will purify the state. This is done by considering the thermofield double of the original CFT mixed state. The resulting geometry is formed by glueing two copies along the black hole horizon resulting in a two sided wormhole geometry which is topologically a cylinder with two boundaries. Now the minimal surface is the throat of the resulting wormhole. We can view the bit threads in two ways (see figure \ref{fig:tfdthreads}):
\begin{outline}
\1 Construct a maximal bit thread configuration on each copy. Since the threads saturate at the black hole horizon we can identify threads when gluing the two copies. This leads to a maximal thread configuration which saturates the wormhole throat.
\1 In the wormhole geometry construct a maximal flow directly. Such a flow will saturate the wormhole throat.
\end{outline}
These give ``constructed" bit thread configurations and ``optimized' bit thread configurations respectively. In general all constructed bit thread configurations can be realized as optimized bit thread configurations; the reverse is not necessarily true.

\subsubsection{Bipartite entanglement wedge cross section}

\begin{figure}[H]
\centering
\begin{tabular}{cc}
\includegraphics[width=.3\textwidth]{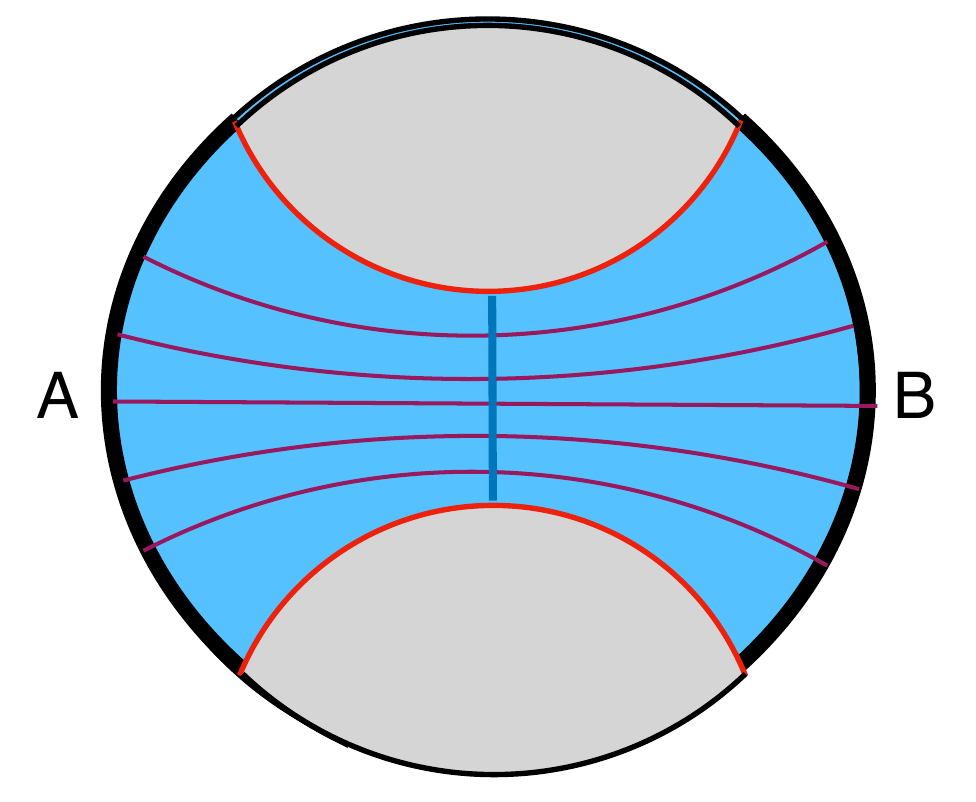} & \includegraphics[width=.3\textwidth]{figs/bibt} \\
$\mathcal{A}$ & $\mathcal{B}^{*}$ \\[6pt]
\multicolumn{2}{c}{\includegraphics[width=.5\textwidth]{figs/tfdthreads} }\\
\multicolumn{2}{c}{$\mathcal{P}$}
\end{tabular}
\caption{\label{fig:biconstructed} We start with a copy of the entanglement wedge $\mathcal{A}$ and place on it the entanglement wedge cross section $b_{p}(A)$ and the maximal flow $v_{A}$. Taking the CPT conjugate we create a second copy $\mathcal{B}^{*}$ and place on it $b_{p}(B)$ and $v_{B} $. In this simple bipartite case we can take  $b_{p}(A) = b_{p}(B)$ and  $v_{A} = v_{B}$. This will not be necessarily true in the multipartite case. Now we can glue $\mathcal{O}$ together identifying the surfaces on $\mathcal{A}$ and $\mathcal{B}^{*}$. This creates the purified manifold $\mathcal{P}$. Using the original cross sections and flows, we can identify the new entanglement wedge cross section and a maximal flow on $\mathcal{P}$. Since these were minimal (maximal) respectively on each copy they too are minimal (maximal) on $\mathcal{P}$. This is our procedure for making ``constructed" flows.}
\end{figure}

\begin{figure}[H]
\centering
\begin{tabular}{cc}
\includegraphics[width=.5\textwidth]{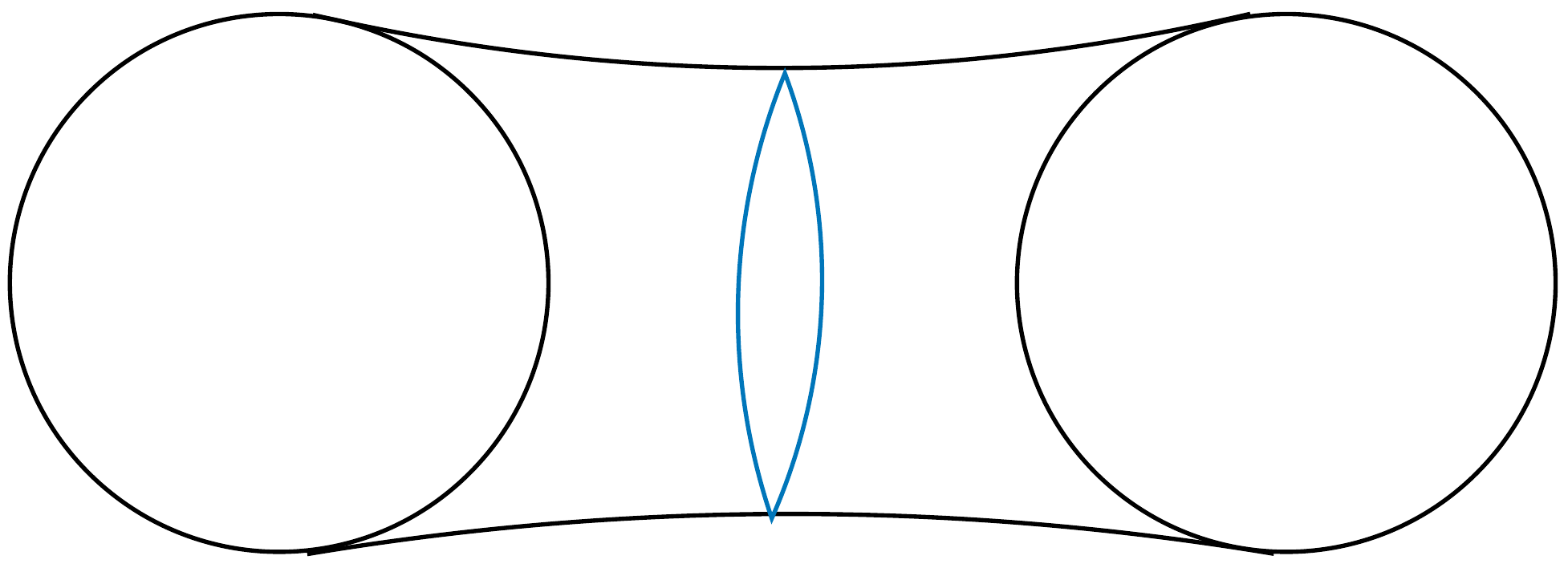} & \includegraphics[width=.5\textwidth]{figs/tfdthreads}
\end{tabular}
\caption{\label{fig:bioptimized} Directly forming $\mathcal{P}$ we can optimize to find the minimal surface and maximal flow. Duality guarantees the flow will saturate on the minimal surface. This is our ``optimized flow". }
\end{figure}

Starting from a two party mixed state we can form the entanglement wedge and calculate the entanglement wedge cross section. First, we purify the geometry by glueing the entanglement wedge along its boundary to its CPT conjugate. This results in a new purified manifold $\mathcal{P}$ which is topologically a cylinder with boundary. One should think of this geometry as dual to a CFT state on the combined boundary \cite{Dutta:aa}. On the purified geometry one can calculate the entanglement entropy between the two sides. Using the RT formula this is given by the area of the wormhole throat. An important feature here was that the purification allowed the entanglement wedge cross section to be related to topological data of the new constructed manifold. We can then create constructed bit thread configurations (figure \ref{fig:biconstructed}) and optimized bit thread configurations (figure \ref{fig:bioptimized}).

These simple examples serves to demonstrate several heuristics which will be helpful as we proceed:
\begin{outline}
\1 Mixed state and multipartite entanglement will often involve bit threads which end on boundaries other than the CFT boundary.
\1 By purifying the geometry these boundaries are eliminated, but the threads can naturally be identified with one another.
\1 This allows for two constructions: 
\2 a construction on patches which are glued together
\2 a direct construction by finding a maximal flow on the purified manifold.
\1 Purification can naturally lead to geometries with non-trivial topology.
\end{outline}

\subsection{Dualization using antiloops}
Here we present an alternative dualization of the standard max flow/min cut program. Usually, a convex relaxation is performed which brings the program into a form which can be easily dualized. A scalar field is used to smear the levels sets of the surface while different boundary conditions are placed on each boundary. This frustrates the system and requires the scalar to interpolate between these two values. When minimizing the problem wants to put this transition at exactly the minimal surface where the least cost is accrued. Thus, the scalar field after minimization is a delta function at the minimal surface. For our purposes, the issue with this procedure is the need to privilege one boundary from another. In the multipartite setting, where we have more than two boundaries, we want a protocol which allows us to place them all on equal footing.
We will instead use a different choice of convex relaxation which treats all of the boundaries equally. The main idea is to use the underlying topology of the manifold to fix the necessary boundary conditions after smearing level sets \cite{Headrick:2018ncs}. This has the advantage that we can refer to the boundary of the manifold as a whole, there is no need to identify an $A$ or $B$ subregion. Later this will become a necessary tool to describe the dual flow program to surfaces in more involved homology classes.

\begin{figure}[H]
\centering
\includegraphics[width=.5\textwidth]{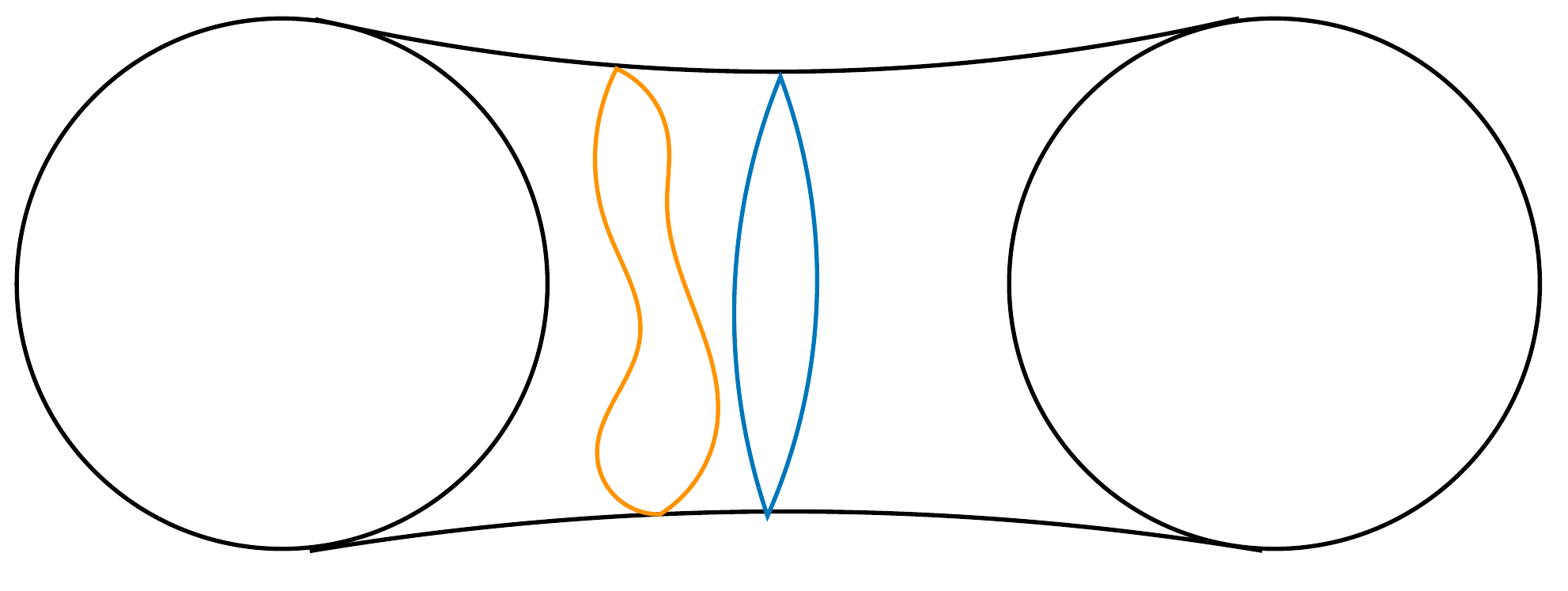}
\caption{\label{fig:tfddefect} Our antiloop $l$ is shown in and orange and the minimal surface in blue. The antiloop give us boundary conditions which allow the minimization program to calculate the minimal surface.}
\end{figure}

To begin, we start with the minimization program for the entanglement entropy between the two sides of the two-sided wormhole geometry dual to the thermofield double
\begin{equation}
S(A)=\min_{m\sim A}\area(m).
\end{equation}
By purity this is also $S(B)$. Now, on this manifold there are two classes of closed simple loops: those with trivial homology which can be contracted to a point, and those with non-trivial homology which wrap the cylinder. Let us call the set of all such non-contractable surfaces $G$. It is clear that the entanglement entropy can also be written
\begin{equation}\label{eq:tfdming}
S(A)=\min_{m\in G}\area(m).
\end{equation}
It is from this this viewpoint that we will perform the convex optimization. We start by smearing the level surfaces and demanding the following boundary conditions and constraints on the resulting scalar field $\psi$: First, we set the boundary to zero that is $\psi\vert_{\partial}=0$. Note if this was our only boundary condition then the solution $\psi=0$ would be feasible. In other words this constraint alone picks out the curves of trivial homology, the smallest of which has area $0$. We need a second constraint which will frustrate the system and force the resulting solution to be an element of $G$. This is accomplished by utilizing an ``antiloop": we pick and arbitrary element of $G$, call it $l$, and demand that crossing the antiloop causes an increase of $\delta\psi\vert_{l}=1$ (see figure \ref{fig:tfddefect}). This step function increase frustrates the system forcing $\psi$ to change elsewhere in the manifold and the best location to do this, which incurs the least cost, is exactly the minimal surface which is homologous to our antiloop $l$\footnote{Though we are illustrating the use of an antiloop for a static 2-dimensional geometry this method straightforwardly generalizes to higher dimension so long as we can find a non-contractible surface in the same homology class as the minimal surface of interest.}.

Performing the relaxation of \ref{eq:tfdming} and imposing the constraints we have the Lagrangian
\begin{equation}
L=\int_{M}\sqrt{g}\vert\nabla_{\mu}\psi-n_{\mu}\delta_{l}\vert+\int_{\partial}\sqrt{h}\alpha\psi+\int_{l}\sqrt{h}\beta(\delta\psi-1)
\end{equation}
where $\alpha$ and $\beta$ are Lagrange multipliers imposing our constraints. Note that in this Lagrangian we have explicitly subtracted the contribution of $\psi$ on $l$, $\delta_l$, from the objective. This is to make sure we only count the change in $\psi$ which is not due to the constraints. After introducing the covector $w_{\mu} = \nabla_{\mu}\psi-n_{\mu}\delta_{l}$ and integrating by parts we now have
\begin{equation}
\int_{M}\sqrt{g}\vert w_{\mu}\vert+v^{\mu}\left(-w_{\mu}+\nabla_{\mu}\psi-n_{\mu}\delta_{l}\right)+\int_{\partial}\sqrt{h}\left(\alpha+n_{\mu}v^{\mu}\right)\psi+\int_{l}\sqrt{h}\beta(\delta\psi-1).
\end{equation}
We are now ready to perform convex dualization by integrating over the original variables $\psi$ and $w_{\mu}$. Doing so gives us the dual maximization program
\begin{equation}
\max_{v^{\mu}}\int_{l}\sqrt{h}v^{\mu}n_{\mu} \quad \text{s.t.} \quad \vert v\vert \leq 1, \; \nabla_{\mu}v^{\mu} = 0.
\end{equation}
Finally, using the divergencelessness of the flow and that $l$ is homologous to the true minimum surface $m$ as well as the boundary $\partial$ we can evaluate the maximal flow on any of these surfaces. This gives us the standard bit thread dual for the entanglement entropy
\begin{equation}
S(A)=\max_{v^{\mu}}\int_{A}\sqrt{h}v^{\mu}n_{\mu} \quad \text{s.t.} \quad \vert v\vert \leq 1, \; \nabla_{\mu}v^{\mu} = 0.
\end{equation}

The use of the antiloop method for this simple example was unnecessary. However, it serves to highlight an important feature of the duality: it was able to be done with reference to only the topological data of the manifold and without reference to the boundary. Later we will return to this method to perform convex duality when the surface of interest is more complex and not in the same homology class as the boundary.

\section{Multipartite entanglement wedge cross section}\label{sec:mreflected}

In this paper we will propose a purification procedure for the multipartite entanglement wedge. For concreteness we will work explicitly with the case of a static slice of AdS$_{3}$ which will allow us to consider simple gluings along geodesics. As we proceed our task will be to use the original entanglement wedge and its CPT conjugate to define a purification by a gluing procedure which eliminates all boundaries which are not a part of the original CFT. We will specifically require that this purification relates the multipartite entanglement wedge cross section to topological data of a newly constructed manifold $\mathcal{P}$. It is this requirement which will act as our guiding principle in the construction.

\subsection{Constructing the purified geometry $\mathcal{P}$}

As a simple, but non-trivial example of our procedure for the purification of the multipartite entanglement wedge we will consider AdS$_{3}/$CFT$_{2}$ with three boundary regions $A,B,C$. Here our static time slice is the hyperbolic disk, $r(ABC)$ is a submanifold, and $\mathcal{O}$ is a union of geodesics. 

Were we to follow the standard procedure for purification we would glue $r(ABC)$ to its CPT conjugate along each piece of $\mathcal{O}$. The resulting manifold would be a ``pair of pants geometry": a Riemannian manifold with genus zero and three boundaries. The minimal surface homologous to $ABC$ is the sum of the wormhole throats with value $2\area(b_{p})$. However, this construction does nothing to map $\Sigma$ to any feature of the manifold. In short this construction is too simple. A more involved procedure is needed to characterize $\Sigma$.

\begin{figure}[H]
\centering
\includegraphics[width=.75\textwidth]{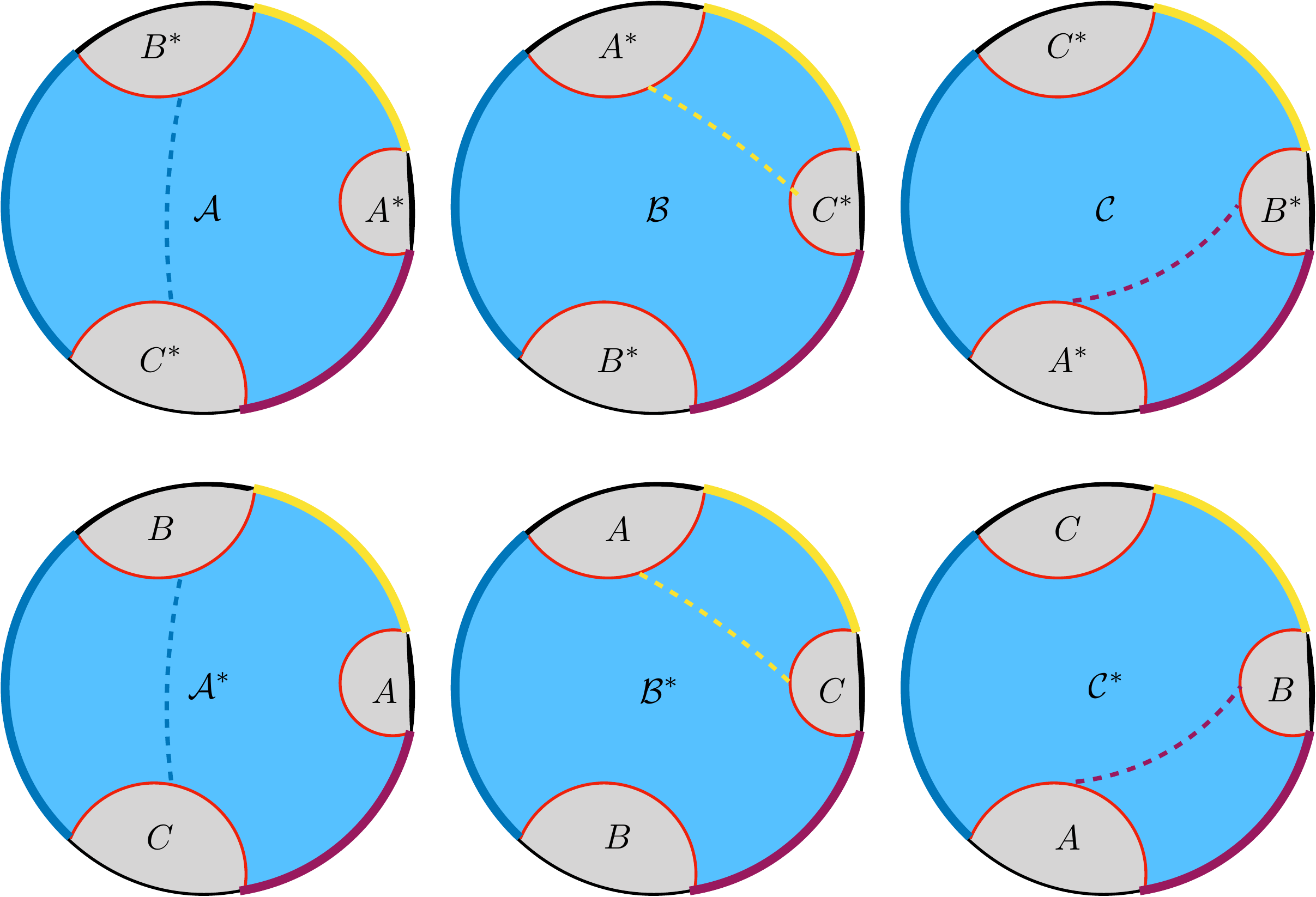}
\caption{\label{fig:6copiesgluing} The 6 copies each labeled in the center. Each gray region labels which copy the surface will be glued to. Note the pieces of $\Sigma$ connect sequentially $\Sigma(A)$ to $\Sigma(B)$ to $\Sigma(C)$ etc. forming a double cover.}
\end{figure}

Instead, we will consider the following more involved construction: Take three copies of $r(ABC)$ and three of its CPT conjugate (since we are going to be gluing this is done so that the total number of surfaces to glue 3*6 is even) and label the copies $\mathcal{A},\mathcal{A}^{*},\mathcal{B},\mathcal{B}^{*},\mathcal{C},\mathcal{C}^{*}$. We then glue the pieces of $\mathcal{O}$ in two steps with the following rules:

\begin{outline}
\1On the copies $\mathcal{A},\mathcal{A}^{*}$ include the surface $\Sigma(A)$. Do the same for $\mathcal{B},\mathcal{B}^{*}$ and $\mathcal{C},\mathcal{C}^{*}$ copies.
\1 Always glue matching copies of the same geodesic piece of $\mathcal{O}$.
\1 Each geodesic on an $\mathcal{A}_{i}$ copy is glued to a geodesic of an $\mathcal{A}^{*}_{j}$ copy. That is the gluings are always between one copy and its CPT conjugate.
\end{outline}
We now perform the first step: Using these rules we glue the pieces of $\Sigma$ so that they form a single closed loop. For example, the two boundaries which $\Sigma(A)$ touches on $\mathcal{A}$ are glued to $\mathcal{B}^{*}$, $\mathcal{C}^{*}$ respectively. Second, the remaining gluings are chosen so that the copies of the three original boundaries $A,B,C$ each form a single oriented boundary in the new geometry. In the case of three boundary regions the above rules automatically guarantee this. Together this procedure completely determines the gluing of all $18$ geodesics (see figure \ref{fig:6copiesgluing}).

\begin{figure}[H]
\centering
\begin{tabular}{c}
\includegraphics[width=.5\textwidth]{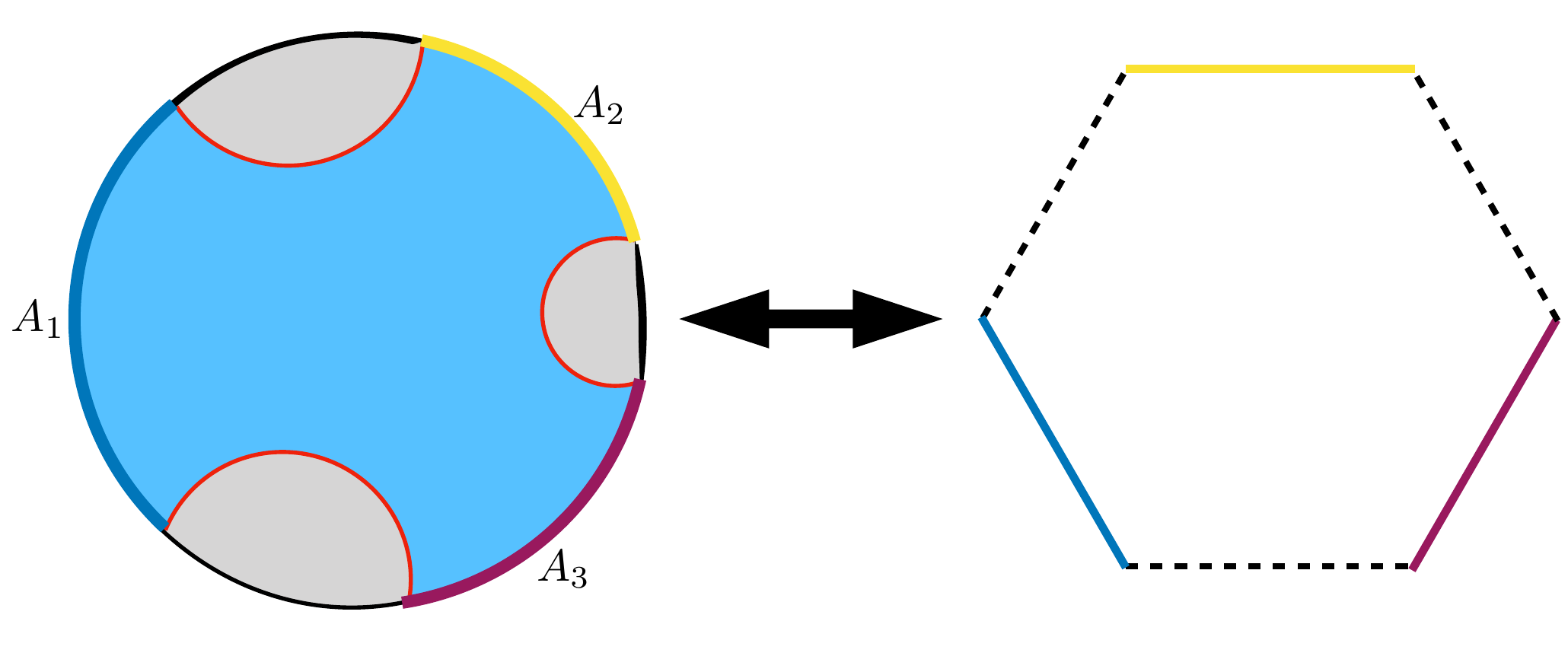}  \\
\includegraphics[width=.7\textwidth]{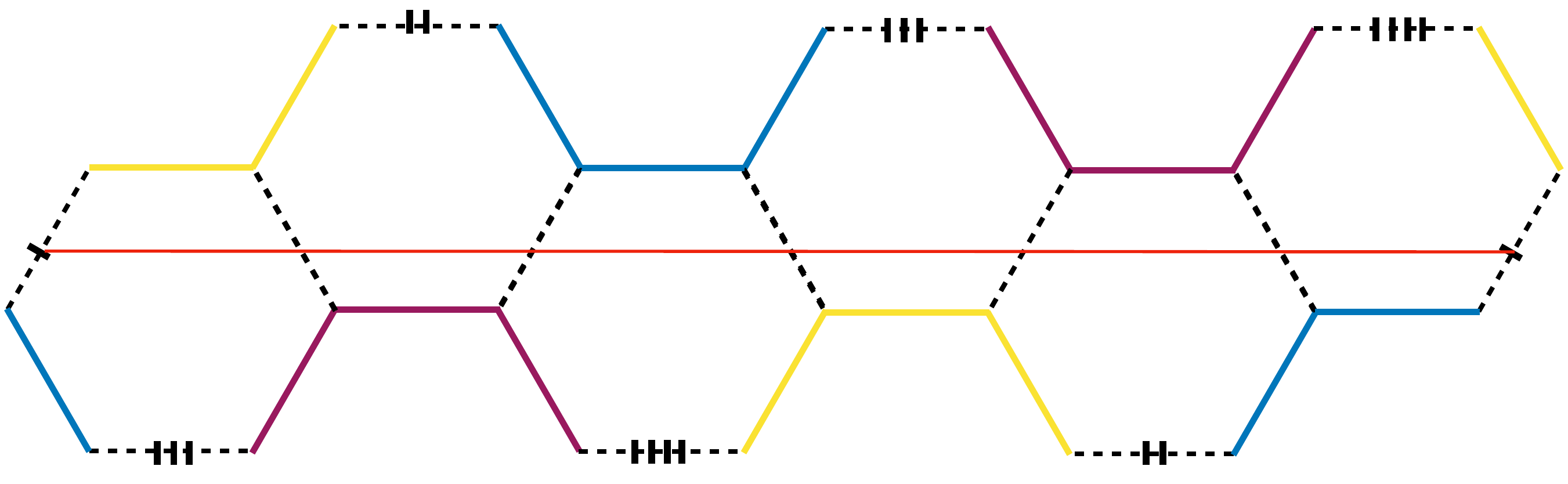} \\
\end{tabular}
\caption{\label{fig:hextorus} Top: We can represent each copy of the entanglement wedge as a hexagon. Bottom: The six copies from figure \ref{fig:6copiesgluing} glued to form the manifold $\mathcal{P}$. The solid red line is the resulting representation of $\Sigma$ on $\mathcal{P}$ here called $\Sigma_{\mathcal{P}}$.}
\end{figure}

Topologically each individual copy can be viewed as a hexagon which after gluing following our prescription gives us a topological representation of the purified manifold $\mathcal{P}$: figure \ref{fig:hextorus}. 

\begin{figure}[H]
\centering
\begin{tabular}{cc}
\includegraphics[width=.25\textwidth]{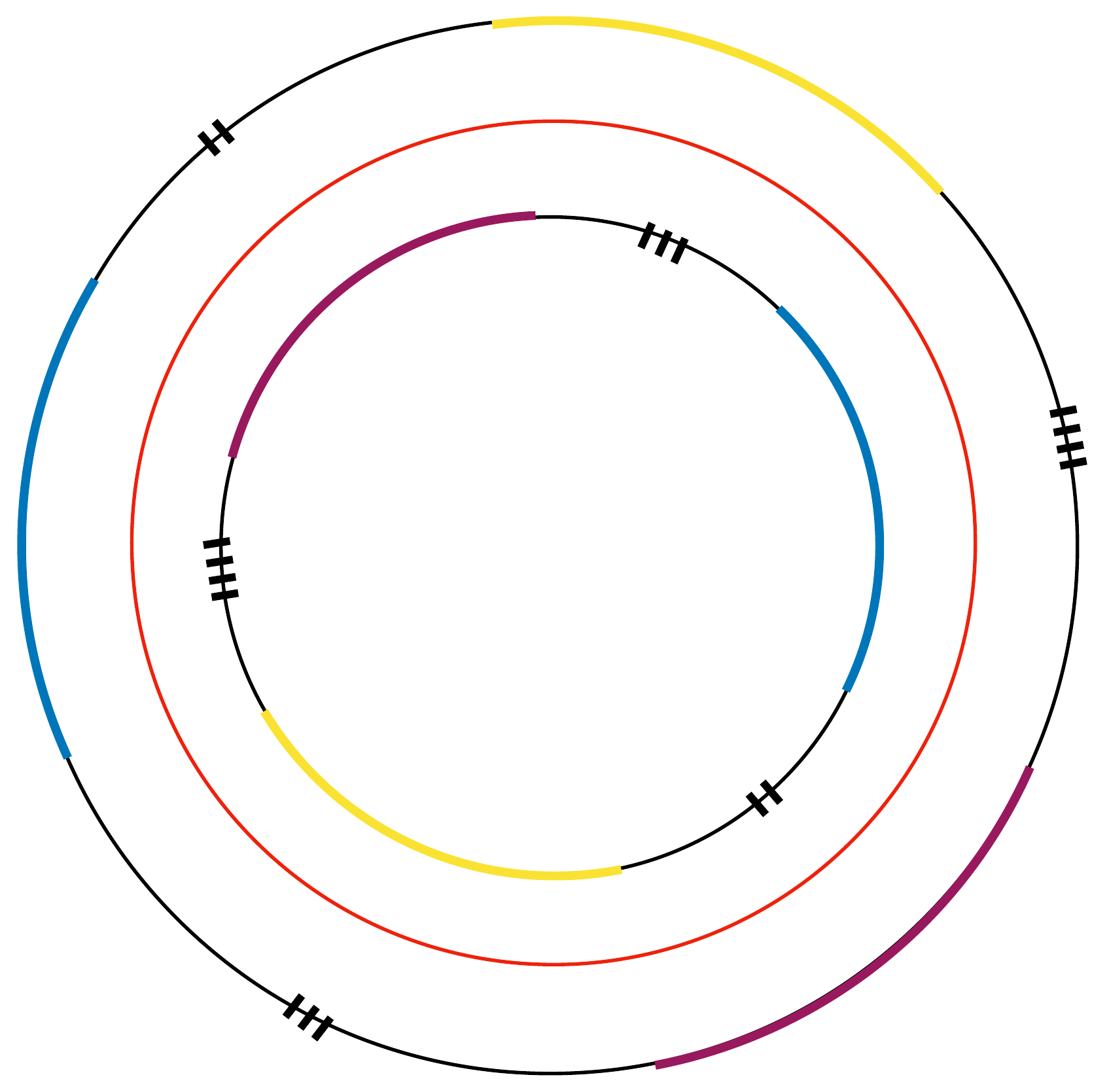} & \includegraphics[width=.25\textwidth]{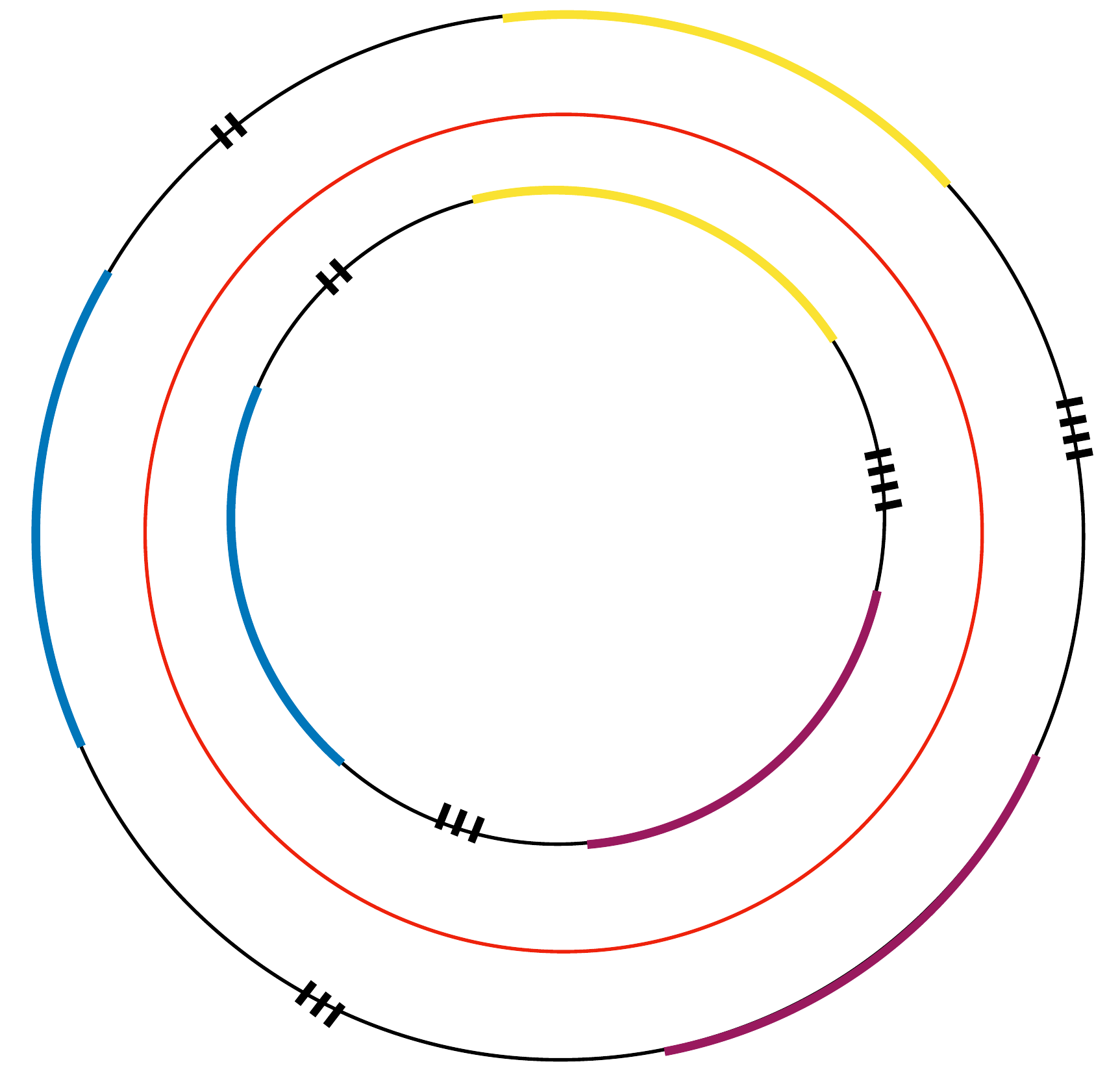} \\
\multicolumn{2}{c}{\includegraphics[width=.26\textwidth]{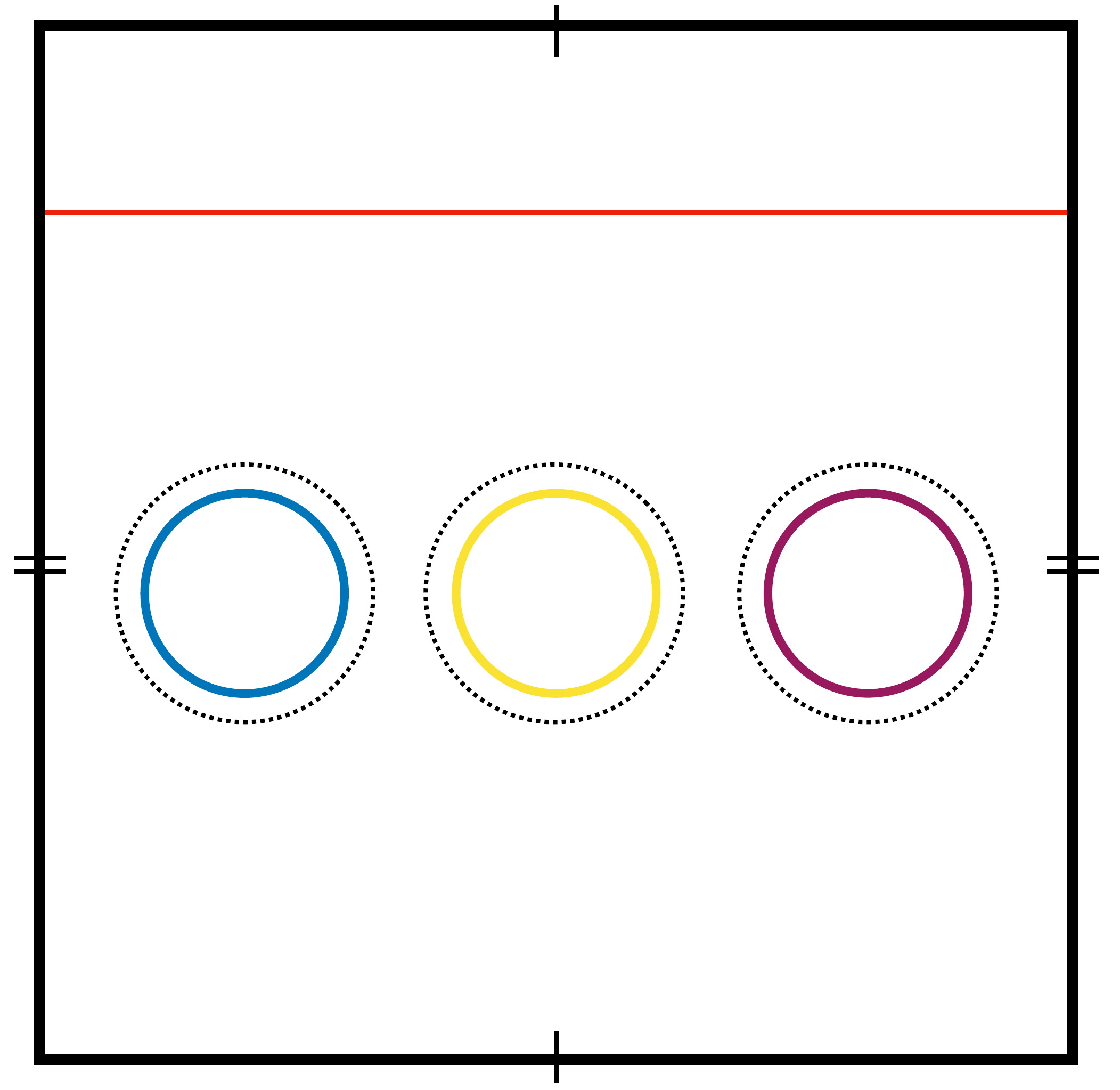} }\\
\end{tabular} 
\caption{\label{fig:annulus} Top left: $\mathcal{P}$ after gluing tab $1$. Top right: The annulus after rotation. Bottom: A topological representation of the manifold $\mathcal{P}$. Shown are the three boundaries, $b_{p\mathcal{P}}$ the dashed lines, and $\Sigma_{\mathcal{P}}$ in red. }
\end{figure}

Our goal now will be to understand the topological properties of $\mathcal{P}$. This can be most easily seen by performing a series of manipulation which leave invariant the topological data. First, we explicitly perform the gluing of the ends of $\mathcal{P}$ giving us a cylinder with identified boundaries. This cylinder can then be mapped to an annulus. We can then rotate the inner and outer boundaries with respect to one another which amounts to mapping straight lines to curved lines. This allows us to line up the remaining surfaces. From this diagram we roll the annulus and perform the last identifications. The resulting manifold after gluing, $\mathcal{P}$, is a torus with three boundaries (see figure \ref{fig:annulus}). Furthermore, we clearly see the representation of the surface $\Sigma$ on $\mathcal{P}$, $\Sigma_{\mathcal{P}}$, has non-trivial homology as it wraps a cycle of the torus.

\subsection{Minimal surfaces on $\mathcal{P}$ and constructed thread configurations}

Having determined the relationship between $r(ABC)$ and $\mathcal{P}$ we can now use this knowledge to ask questions about minimal surfaces and maximal flows on $\mathcal{P}$. Importantly, a maximal bit thread configuration for a particular optimization problem will give rise to a constructed bit thread configuration on $\mathcal{P}$. 

\begin{figure}[H]
\centering
\includegraphics[width=.75\textwidth]{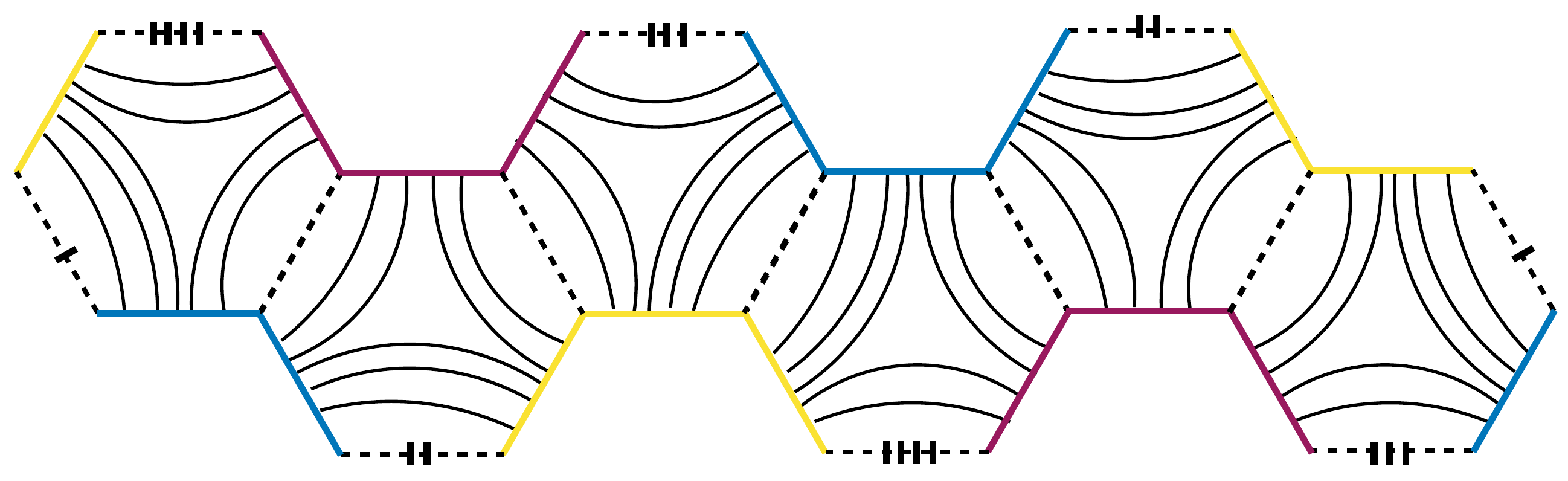}
\caption{\label{fig:bpPconstructed} Using the flow from figure \ref{fig:bp} we can create a constructed flow on $\mathcal{P}$ which saturates on $b_{p\mathcal{P}}$.}
\end{figure}

To start consider a max multiflow on each copy of $r(ABC)$; we know that this is dual to and computes the area of $b_{p}$. Considering each of these six flows together generates a max multiflow on $\mathcal{P}$ (see figure \ref{fig:bpPconstructed}). Since each piece of $b_{p}$ is saturated on each copy this implies $b_{p\mathcal{P}}$ too is saturated. From our analysis of the topology this surface is the sum of the independent minimal surfaces homologous to each boundary region of $\mathcal{P}$. That is the ``throats" of each wormhole. We have found:
\begin{equation}
\sum_{i} \min_{m\sim A_{i} \text{ on } \mathcal{P}}\area{m} = \area(b_{p\mathcal{P}})= 6\area(b_{p}) \leftrightarrow \text{Max multiflow on $\mathcal{P}$}.
\end{equation}

\begin{figure}[H]
\centering
\includegraphics[width=.75\textwidth]{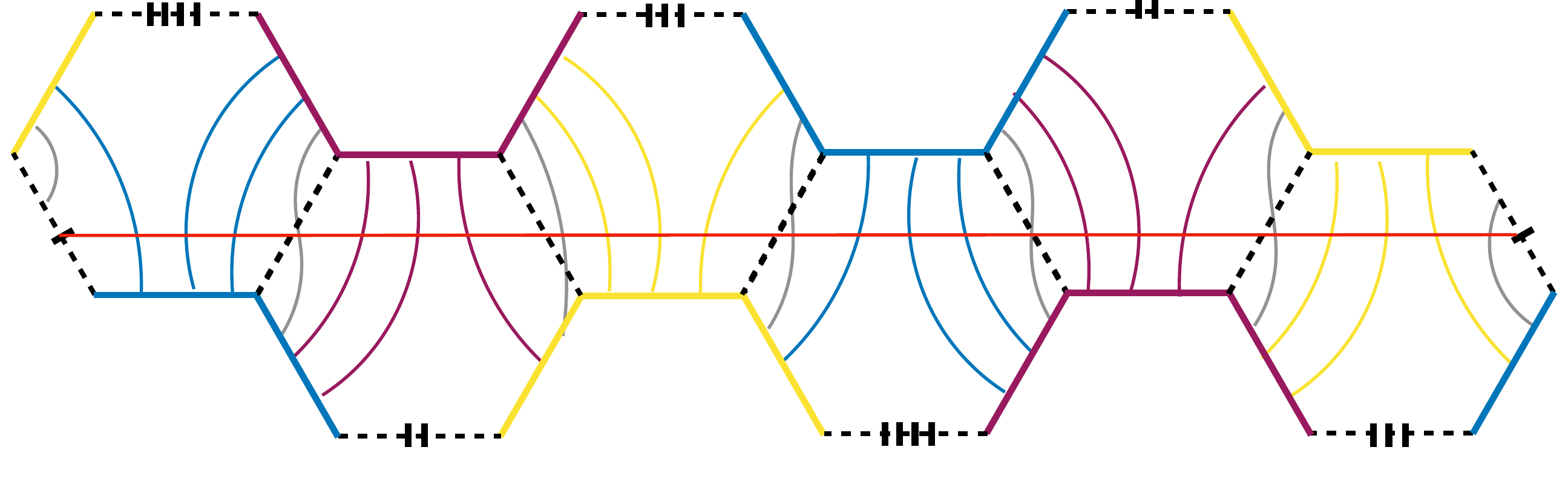}
\caption{\label{fig:tbtconstructed} Using the flows from figure \ref{fig:meopexample} we can create a constructed flow on $\mathcal{P}$ which saturates on $\Sigma_{\mathcal{P}}$. The threads which began on $\mathcal{O}$ (shown in grey) can be identified across the boundaries because of the original constraint $\alpha = n_{\mu}v^{\mu}_{A_{i}}|_{\mathcal{O}}$. It is these threads which cross between the original copies of the geometry which should be thought of as representing the entanglement which is ``truly" multipartite. Note because of how this flow was constructed there are no loops of threads and no threads connecting a boundary of $\mathcal{P}$ back to the same boundary (for example a thread starting and ending on $A$). This guarantees that this flow is in fact a multiflow. This is important as it shows there are maximal multiflows in the space of feasible points of the maximization program used to find optimized flows on $\mathcal{P}$.}
\end{figure}

We can now perform the same procedure using the maximal flows dual to the multipartite entanglement wedge cross section (see figure \ref{fig:tbtconstructed}). On each $A_{i}$ copy of $r(ABC)$ we place the corresponding flow $v_{A_{i}}$. Because of the boundary constraint $n_{\mu}v^{\mu}_{i}|_{\mathcal{O}}=\alpha$ wherever two copies of $r(ABC)$ are glued we can identify the threads on the boundary because the flux is the same. Thus, we will naturally have threads which cross between the different copies of the original geometry $r(ABC)$. Furthermore, we know $v_{A_{i}}$ saturates on its corresponding piece of $\Sigma$, $\Sigma(A_{i})$, so on $\mathcal{P}$ the full surface $\Sigma_{\mathcal{P}}$ is saturated as well. This is only possible if it is the minimal surface in its homology class. From our topological analysis we know this surface is in the homology class of a non-trivial cycle of $\mathcal{P}$. Let $\sigma$ be the set of all such surfaces, then on $\mathcal{P}$ we have:

\begin{equation}\label{eq:sigmatmin}
\min_{m\in \sigma\text{ on }\mathcal{P}}\area{m} = \area(\Sigma_{\mathcal{P}})= 2\area(\Sigma) \leftrightarrow \text{Max multiflow saturating elements of $\sigma$ on $\mathcal{P}.$}
\end{equation}

\subsection{Optimized bit thread configurations on $\mathcal{P}$}
In this section we will perform a convex dualization of the minimization program for $\Sigma_{\mathcal{P}}$ to directly compute its bit thread dual. To do so we will make use of the antiloop technology discussed previously. To begin we use our knowledge that $\Sigma_{\mathcal{P}}$ is the minimal area surface in its homology class $\sigma$. This is purely a topological statement made evident by our construction of $\mathcal{P}$ above
\begin{equation}\label{eq:torusming}
\Sigma_{\mathcal{P}}=\min_{m\in \sigma}\area(m).
\end{equation}

\begin{figure}[H]
\centering
\includegraphics[width=.5\textwidth]{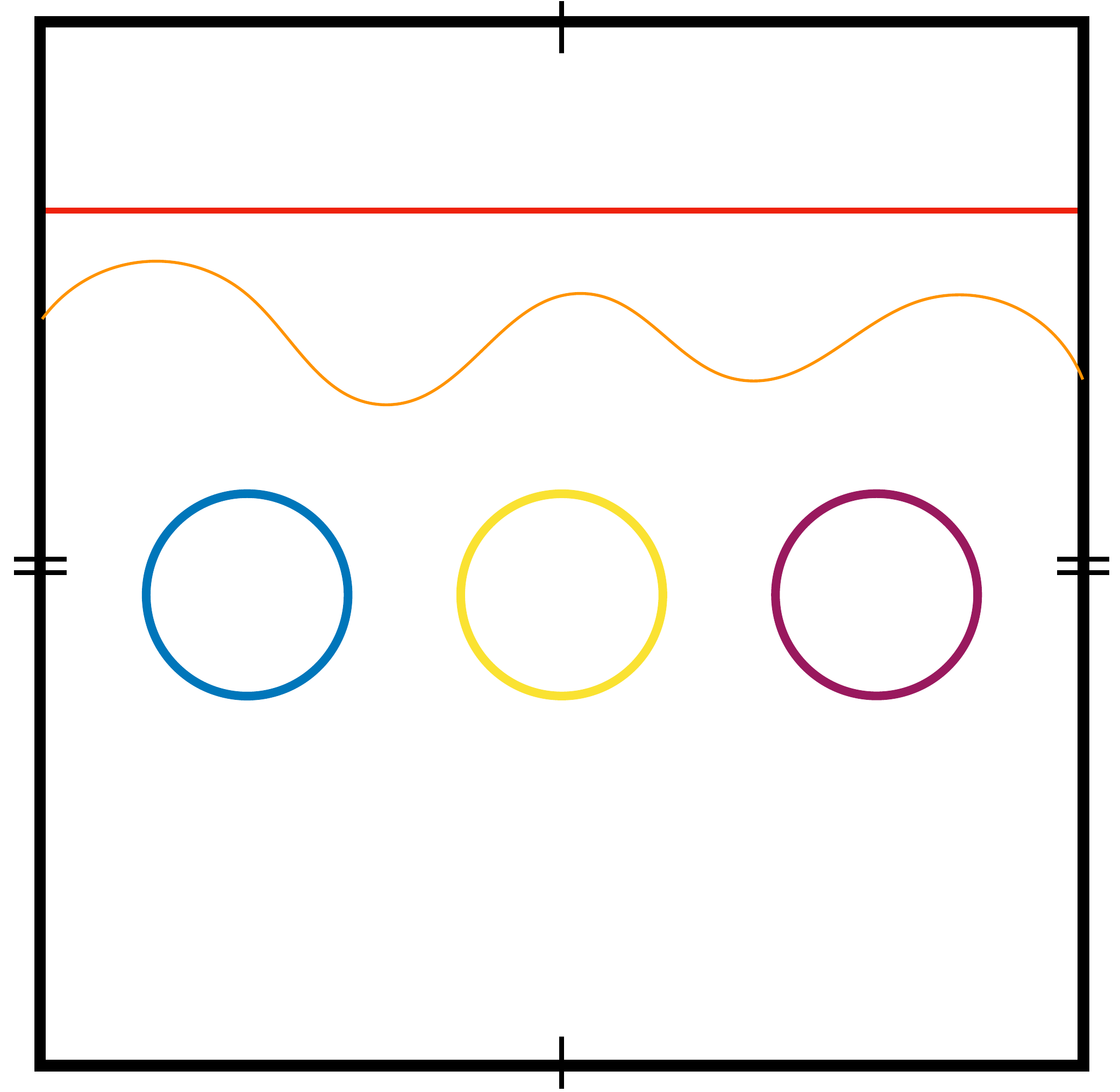}
\caption{\label{fig:torussquaredefect} Picking an element of $\sigma$ as our antiloop $l$ we can perform a direct dualization on $\mathcal{P}$ of the minimization program for $\Sigma_{\mathcal{P}}$.}
\end{figure}
It is important to note that because of the topology of $\mathcal{P}$ we cannot use an element of $\sigma$ to separate the boundary regions. As such it is impossible to impose the normal convex relaxation, this is where the use of the antiloop becomes important. We choose a representative element of $\sigma$, $l$ as our antiloop (see figure \ref{fig:torussquaredefect}). Performing the relaxation of \ref{eq:torusming} and imposing the constraints the dualization proceeds as before
\begin{equation}
l=\int_{\mathcal{P}}\sqrt{g}\vert\nabla_{\mu}\psi-n_{\mu}\delta_{l}\vert+\int_{\partial}\sqrt{h}\alpha\psi+\int_{l}\sqrt{h}\beta(\delta\psi-1).
\end{equation}
where $\alpha$ and $\beta$ are Lagrange multipliers imposing our constraints. Once again in this Lagrangian we have explicitly subtracted the contribution of $\psi$ on $l$, $\delta_l$ from the objective. This is to make sure we only count the change in $\psi$ which is not due to the constraints. After introducing the covector $w_{\mu} = \nabla_{\mu}\psi-n_{\mu}\delta_{l}$ and integrating by parts we now have
\begin{equation}
\int_{\mathcal{P}}\sqrt{g}\vert w_{\mu}\vert+v^{\mu}\left(-w_{\mu}+\nabla_{\mu}\psi-n_{\mu}\delta_{l}\right)+\int_{\partial}\sqrt{h}\left(\alpha+n_{\mu}v^{\mu}\right)\psi+\int_{l}\sqrt{h}\beta(\delta\psi-1).
\end{equation}
We are now ready to perform convex dualization by integrating over the original variables $\psi$ and $w_{\mu}$. Doing so gives us the dual maximization program
\begin{equation}
\max_{v^{\mu}}\int_{l}\sqrt{h}v^{\mu}n_{\mu} \quad \text{s.t.} \quad \vert v\vert \leq 1, \; \nabla_{\mu}v^{\mu} = 0.
\end{equation}
That is our dual flow program is one which maximizes the flux on $l$. Note this program makes no reference to the boundary and in fact can contain loops of threads which wrap the manifold and are not sourced from the boundary\footnote{This is an example of an optimized thread configuration which can not be realized as a constructed one.} To make a connection to the boundary region it is useful to impose some additional constraints on the flow program. This is similar in spirit to the restrictions we made on \ref{meopintro}. Using this original flow program on $r(ABC)$ we know that there exist flows which saturate $\Sigma_{\mathcal{P}}$, which contain no loops of threads, and which contain no threads that begin and end on the same piece of the boundary (e.g. a thread starting on $A$ and traveling back to also end on $A$). This means it is always possible to choose a maximal flow which saturates on $\Sigma_{\mathcal{P}}$ to be a multiflow.

\subsection{The degenerate case}\label{sec:degen}
\begin{figure}[H]
\centering
\includegraphics[width=.3\textwidth]{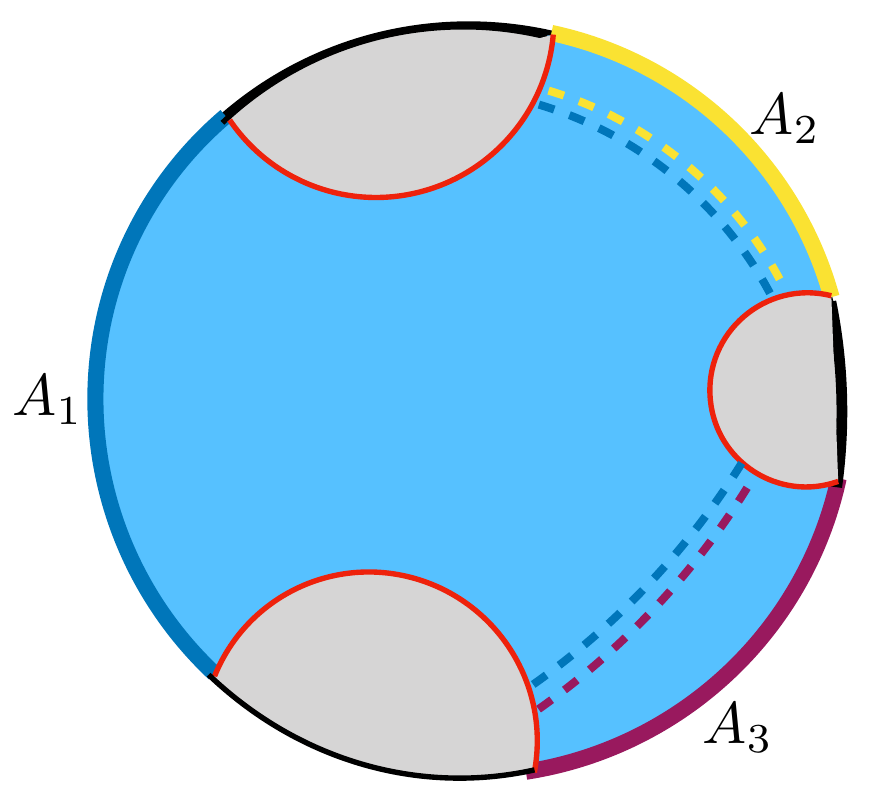}
\caption{\label{fig:upperboundmeop} $\Sigma$ when it obtains its upper bound is built from bipartite entanglement wedge cross sections.}
\end{figure}

We will now consider a special case of the multipartite entanglement wedge cross section when it obtains it upper bound. This occurs when one of the boundary regions is much larger than the others. The minimal surface becomes a sum over the bipartite surfaces excluding the largest (see figure \ref{fig:upperboundmeop}). That is generically
\begin{equation}
\area(\Sigma) \leq 2\min_{i}\sum_{j\neq i}\area(b_{p}(A_{j})).
\end{equation}
This is easy to understand from the bit thread perspective. When one of the boundary regions is large it is able to support enough flux to saturate the individual minimum surfaces for each other boundary region. Importantly, since every thread connects two boundary regions there is no flux needed on $\mathcal{O}$.

\begin{figure}[H]
\centering
\begin{tabular}{cc}
\includegraphics[width=.45\textwidth]{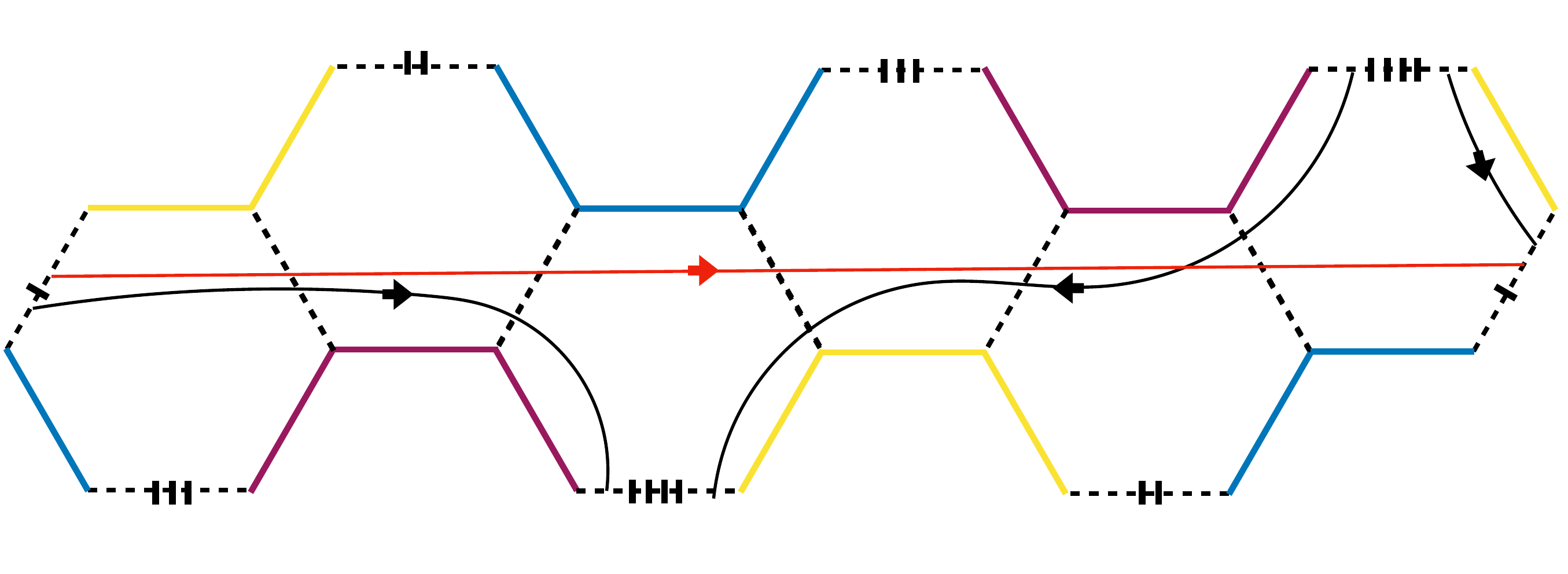} & \includegraphics[width=.45\textwidth]{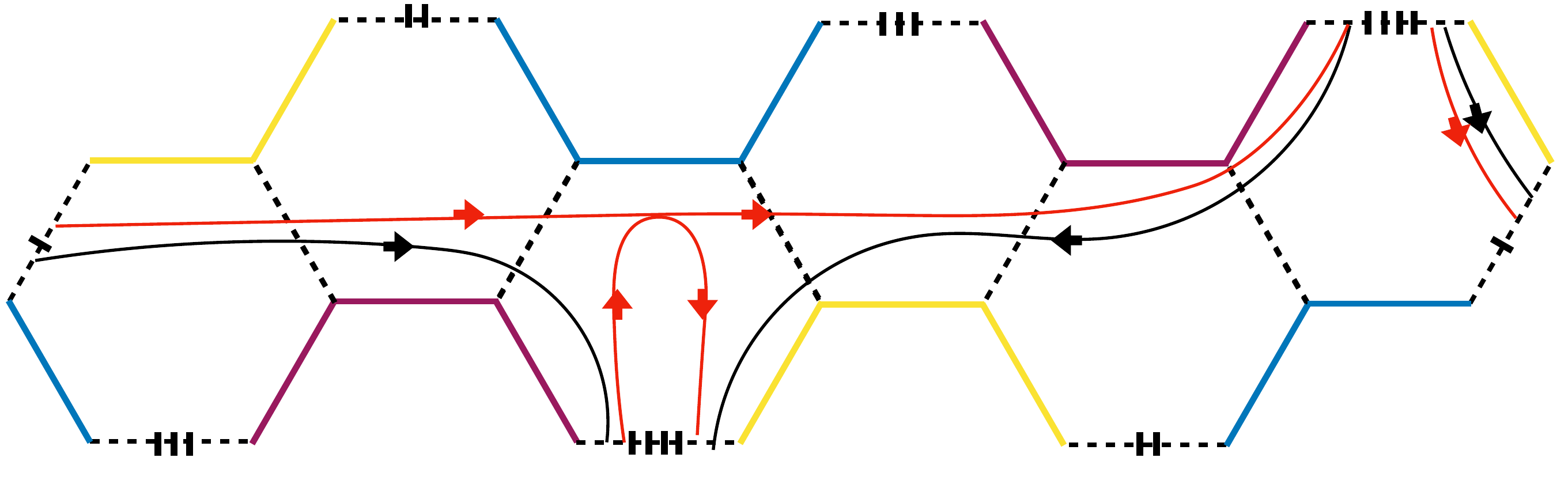} \\
$1$ & $2$ \\[6pt]
\includegraphics[width=.45\textwidth]{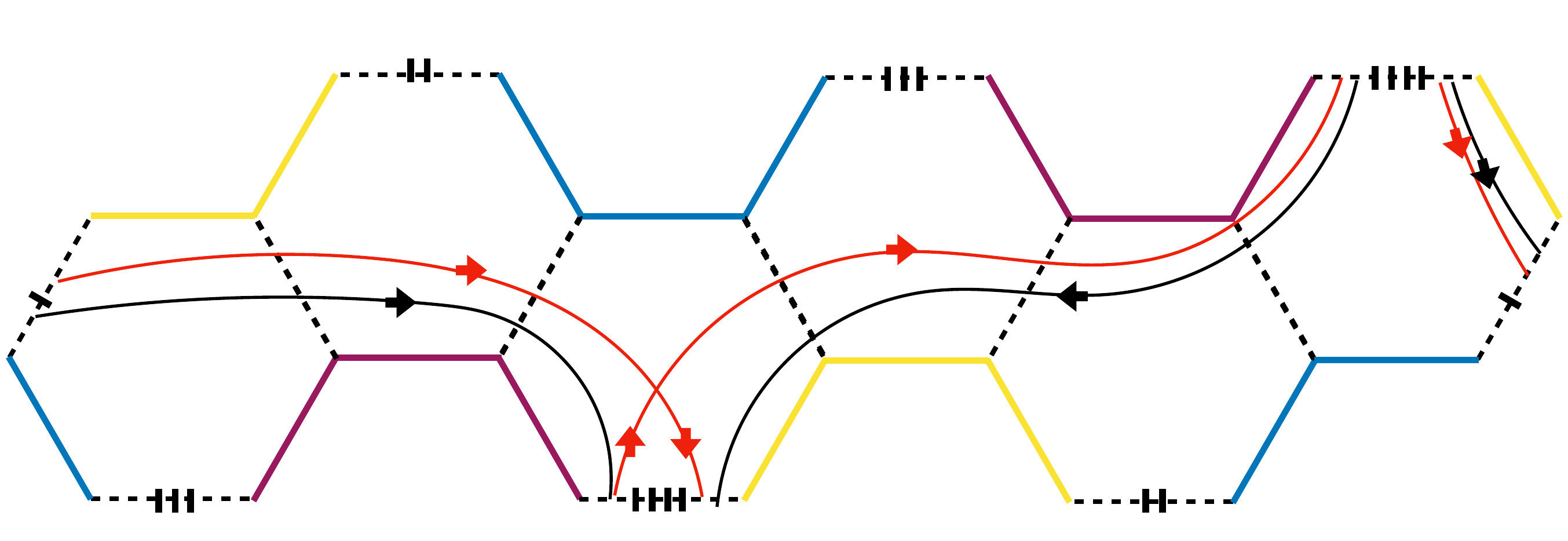} & \includegraphics[width=.45\textwidth]{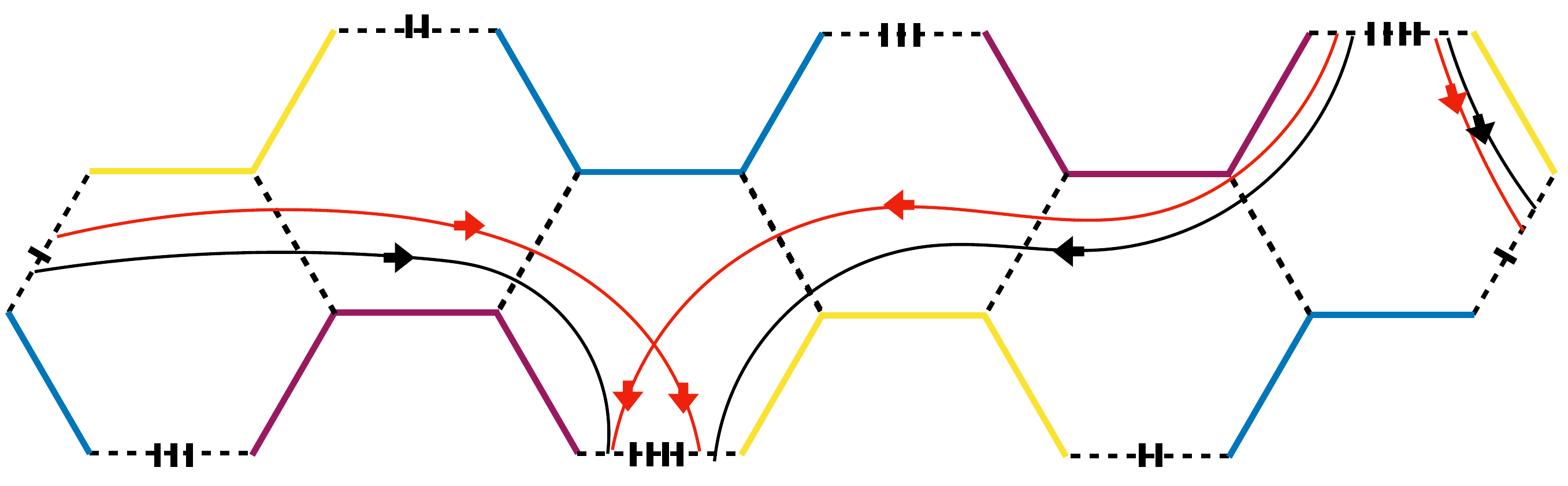} \\
$3$ & $4$ \\[6pt]
\multicolumn{2}{c}{\includegraphics[width=.45\textwidth]{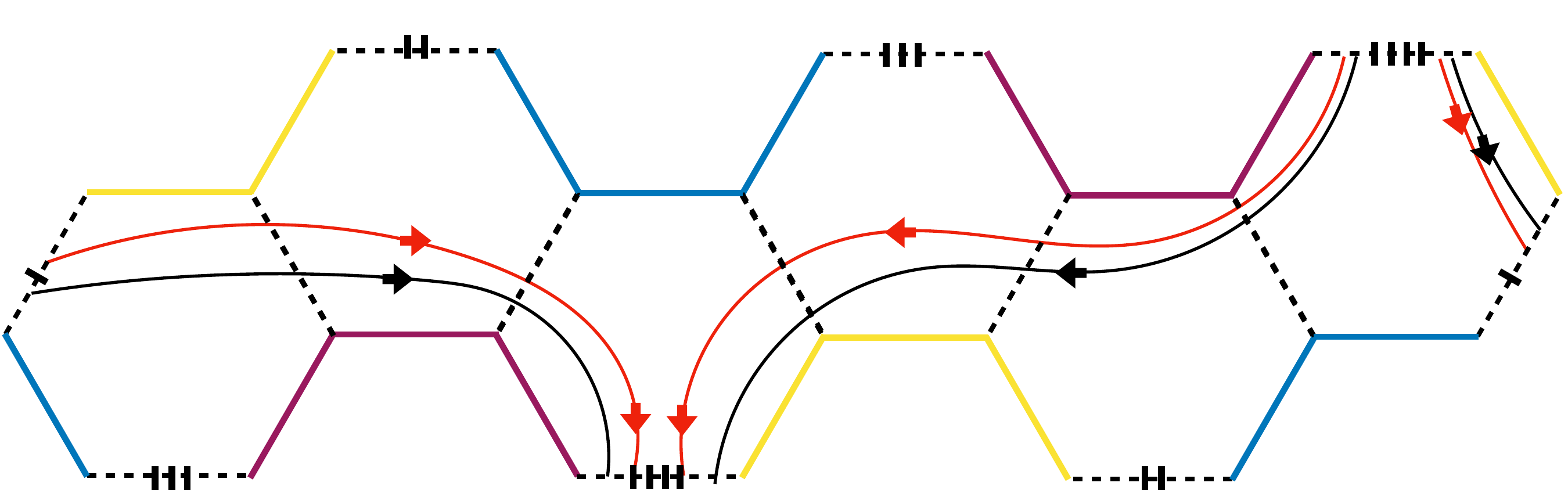} }\\
\multicolumn{2}{c}{$5$}\\
\multicolumn{2}{c}{\includegraphics[width=.3\textwidth]{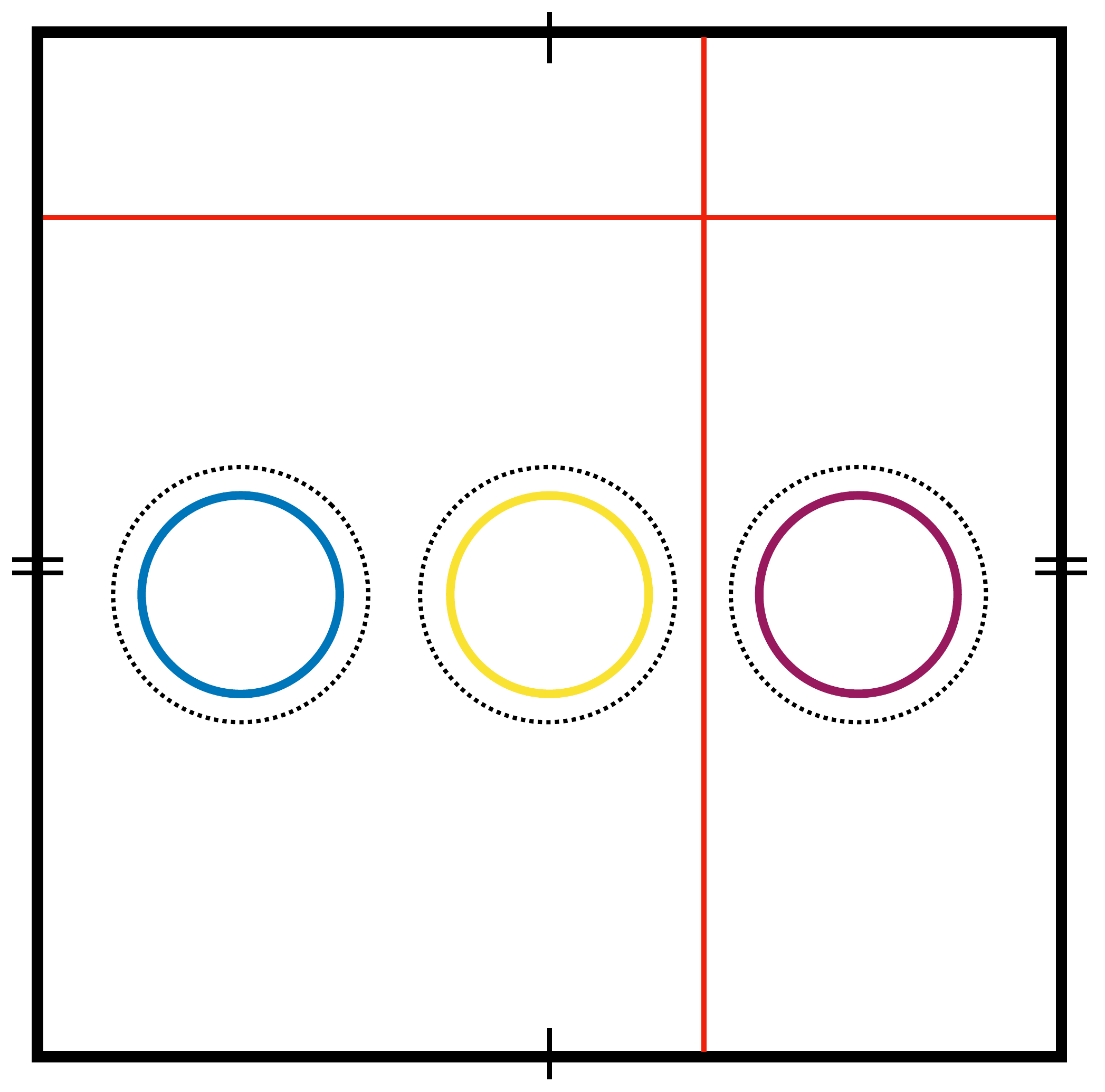} }
\end{tabular}
\caption{\label{fig:degen} 1: Starting with $\Sigma_{\mathcal{P}}$ shown in red on the copies $\mathcal{A}$ and $\mathcal{A}^{*}$ we replace $\Sigma_{A}$ with $b_{p}(B)\cup b_{p}(C)$ to get the surface in black. 2: We push the red surface up so that it touches itself. 3: At the intersection we can then reconnect the surfaces. 4: Acting with the cycle between $B$ and $C$ we reverse the orientation of the middle segment. 5: We can then reconnect once again. The two surfaces are now equivalent.  Bottom:  This procedure shows that the black surface is a combination of two homology cycles. The one which $\Sigma_{\mathcal{P}}$ belongs to and the one between $B$ and $C$.}
\end{figure}
We would now like to repeat the analysis done above for this degenerate case. Using a series of manipulations we can relate the surface $\Sigma_{\mathcal{P}}$ in the usual case with it in the degenerate case: The two surfaces can be related by the action of an additional homology cycle which means the two surfaces are not homologous (see figure \ref{fig:degen}). From the bit thread perspective this phase transition is accompanied by a change in the behavior of the threads. No longer are threads needed which cross between the original copies. Instead, we can use the same maximal multiflow on each copy which saturates the bipartite entanglement wedge cross sections. We can think of this change as a statement about the allowed distillations of the entanglement of the multipartite entanglement wedge cross section. The change in homology class can thus be seen as indicating a loss of ``truly" multipartite entanglement. This illustrates a possible deeper connection between entanglement structure and topology.

\subsection{More boundary regions}
So far we have restricted ourselves to three boundaries, we will now consider more. As the number of boundary regions increases so does the number of copies needed for the purification. For an even number of boundary regions $n$, $n$ copies are needed. $\frac{n}{2}$ copies of the original manifold and $\frac{n}{2}$ of the complex conjugate. These are arranged to alternate ($\mathcal{A}$, $\mathcal{B}^{*}$,$\mathcal{C}$, etc.). When $n$ is odd we need $2n$ copies to a guarantee an even number of boundaries to glue. We take one copy and one CPT conjugate for each boundary region ($\mathcal{A}$,$\mathcal{A}^{*}$, etc.). As $n$ increases the number of gluings needed to be performed increases as well. As we will see in the five boundary case this increase in boundaries leads to multiple valid gluings where $\Sigma_{\mathcal{P}}$ is a minimal surface of non-trivial homology on the purification. Regardless of this ambiguity, any of these purifications map the multipartite entanglement wedge cross section to topological data in the purification.

\paragraph{Four boundary regions}
\begin{figure}[H]
\centering
\includegraphics[width=.5\textwidth]{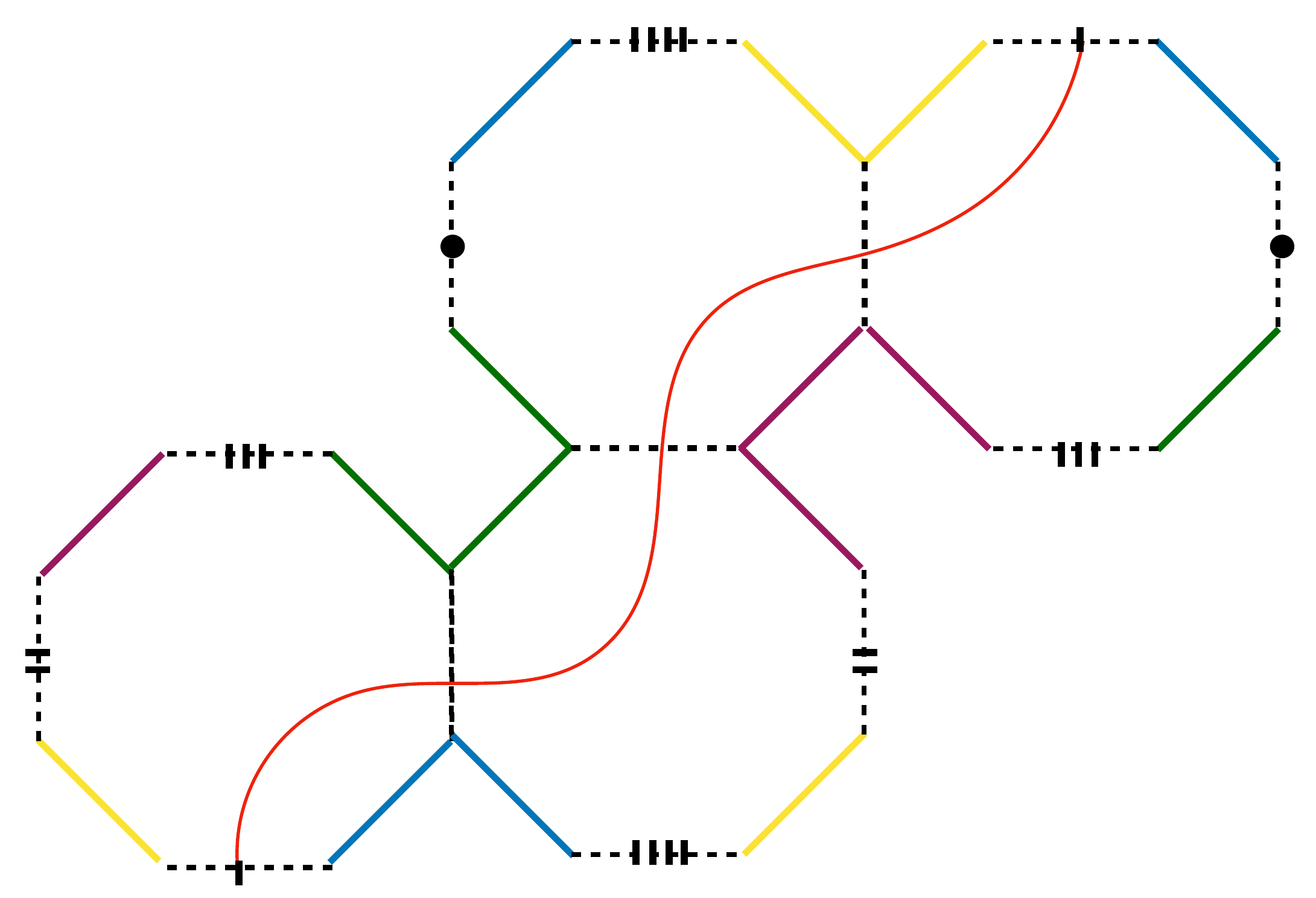}
\caption{\label{fig:octogon}Our construction with four boundary regions using four copies. Because so few gluings need to be performed the requirement that only copies and their CPT conjugate be glued automatically determines all of the identifications. Performing the gluing of tab 1 we can once again rotate the two boundaries and perform the remaining identifications to form a torus with four boundaries.}
\end{figure}
We start with four boundary regions and construct the entanglement wedge which is topologically an octagon. Because we have an even number of boundaries we only need four copies of the geometry. Labeling the copies $\mathcal{A},\mathcal{B}^{*},\mathcal{C} $ and $\mathcal{D}^{*}$ we construct $\Sigma_{\mathcal{P}}$ as before by combining the pieces of $\Sigma(A_{i})$ on each copy to form a closed loop. All that remains is to finish the identifications of the remaining boundaries. In this case all gluing are immediately determined by demanding only identical geodesics on copies and CPT conjugates be identified (see figure \ref{fig:octogon}). The resulting manifold is a torus with four boundaries. 

\paragraph{Five boundary regions}
For five boundary regions we are once again working with an odd number. We take five copies of the original entanglement wedge and five CPT conjugates and glue them cyclically as in the case of three boundary regions so that $\Sigma_{\mathcal{P}}$ forms a closed loop. Now because of the increased number of gluings that must done there is an ambiguity in how to proceed. Here we present two resolutions: 
\begin{figure}[H]
\centering
\includegraphics[width=.7\textwidth]{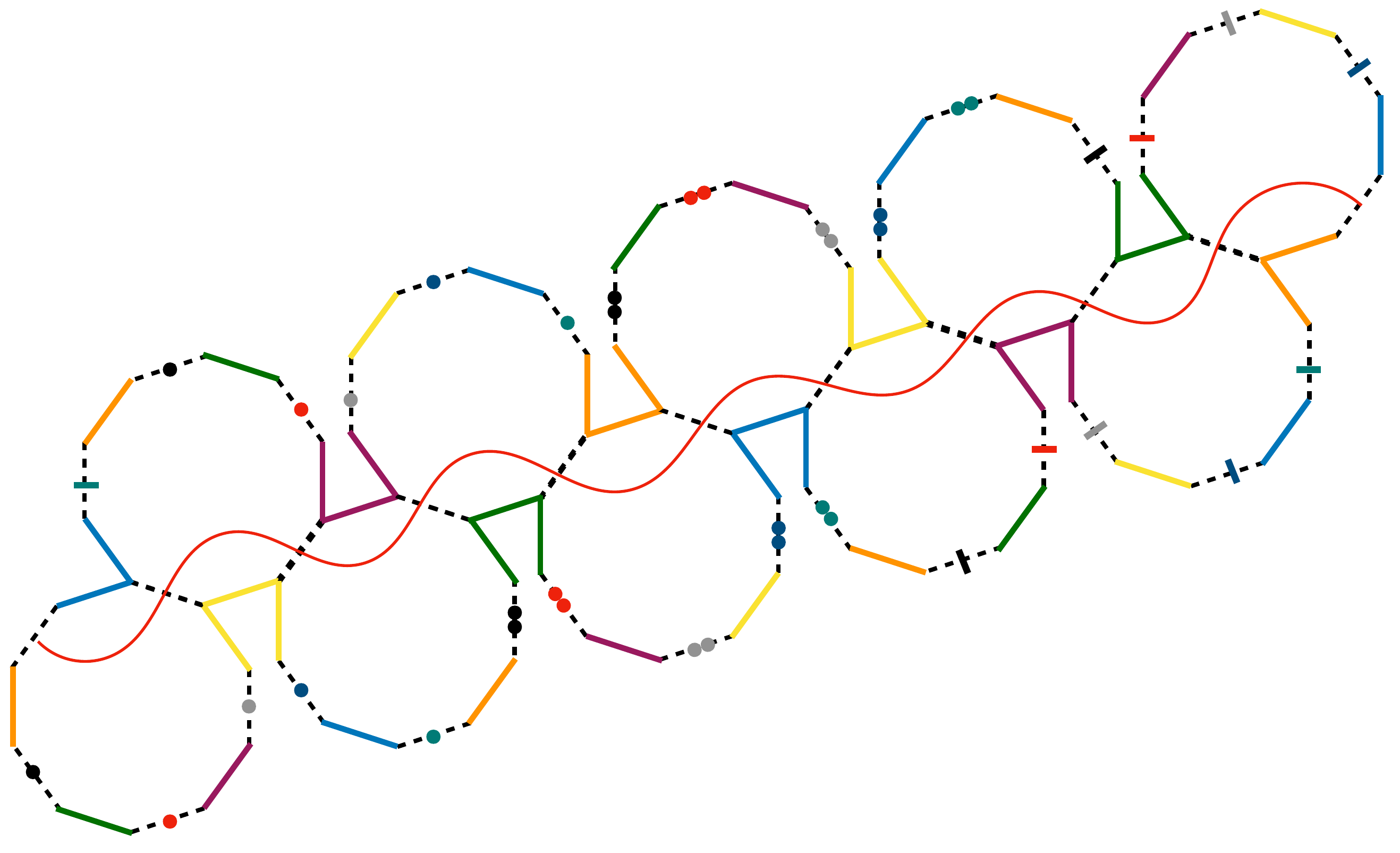}
\caption{\label{fig:decagon1}A choice of gluing for five boundary regions. The geometry is that of a torus, but now with many disconnected boundary regions}
\end{figure}
We can choose for the topology of the purification to be that of a torus. To do so we perform the gluings by identifying surfaces across $\Sigma_{\mathcal{P}}$ offset by two regions (see figure \ref{fig:decagon1}). Incidentally, this was also true in the case of 3 and 4 boundary regions.  In doing so however, we must give up that each of the original boundaries now forms a single oriented boundary in the purification. Instead each boundary will be split into several components.

\begin{figure}[H]
\centering
\includegraphics[width=.7\textwidth]{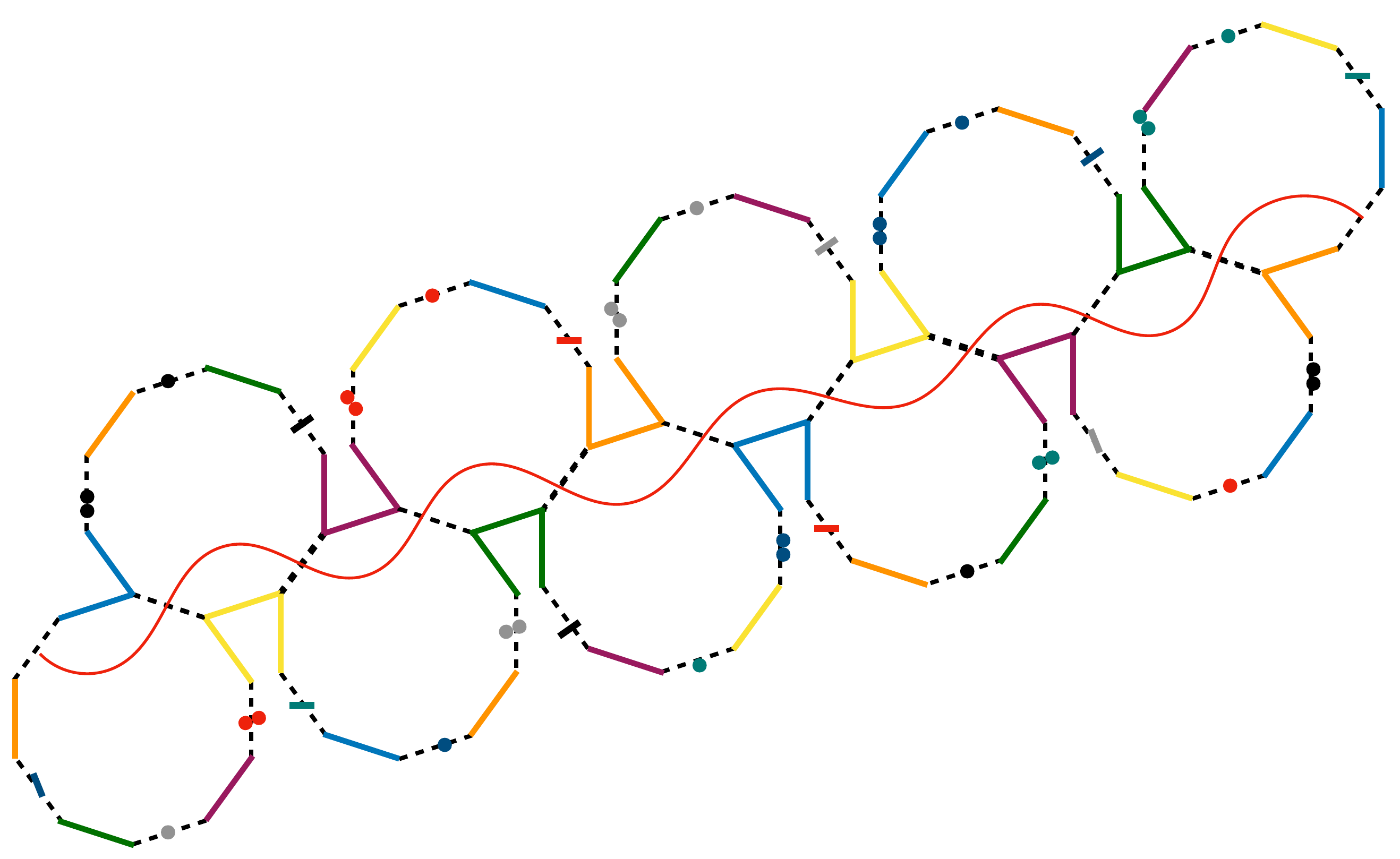}
\caption{\label{fig:decagon2}A choice of gluing for five boundary regions. The topology is no longer that of a torus.}
\end{figure}
Alternatively, we can demand that each copy is glued to its CPT conjugate along the boundary opposite $\Sigma(A_{i})$ the rest of the glueings then immediately follow (see figure \ref{fig:decagon2}). This choice has the advantage of keeping each boundary region as a single oriented loop in the geometry with the tradeoff being the geometry is no longer that of a torus and has more complicated topology. 

Regardless of the choice of gluing used, $\Sigma_{\mathcal{P}}$ retains its interpretation as a closed non-contractible loop which is minimal in its homology class. We expect that this construction can be extended further to an even larger number of boundary regions, but the topology of the resulting manifold will undoubtedly become more complicated.

For more than three boundary regions the degenerate case \ref{sec:degen} becomes harder to analyze. This is because the surface of interest will not necessarily form a single closed loop. This might be possible to accomplish by considering more involved purification procedures.

\subsection{Covariant and higher dimensions}
In order for our method to be more widely applicable we would like to briefly discuss the case of a full covariant spacetime in any number of dimensions. While this may be done, it is important to note the current lack of understanding of the entanglement wedge cross section, especially the multipartite case, in this setting. As such, though we may generalize the gluing procedure there can not be a clear information theoretic-interpretation without these prerequisite definitions and framework in place.

In such a situation we must be able to perform gluings along codimension 2 surfaces. Such a procedure was considered in \cite{Engelhardt:2018kcs} where it was found a consistent gluing between codimension 2 surfaces was possible only if

\begin{equation}
\left[\theta_{k}\right] = 0, \quad \left[\theta_{l}\right] = 0, \quad  \left[\gamma_{i}\right] = 0
\end{equation}
where these are respectively the difference in the expansions and twist potential of the two surfaces to be glued. In our case of interest the surfaces we wish to glue are a copy of an extremal surface and its reflection. Since both are extremal each expansion is identically zero and as a result the difference in expansions, the first two criteria, are always zero. In contrast the twist potential is not guaranteed to be zero, but is invariant under CPT conjugation. Thus each copy of the manifold should only be glued to its CPT conjugate. Doing so ensures the junction conditions are satisfied and a well defined spacetime can be constructed. The metric resulting from this procedure will be smooth, but discontinuities at the junction surface may cause impulsive shockwaves. Following \cite{Dutta:aa} we do not view this as an issue for our construction. This gluing procedure was further formalized in \cite{Marolf:2019zoo} where it was applied to entanglement wedges glued across a common RT surface.

In higher dimensions one must be careful that the choice of boundary regions gives a well defined connected entanglement wedge on which our procedure can be performed. It would be interesting to detail this more fully.

\section{Discussion}\label{sec:discuss}
\paragraph{Reflected entropy}
Several possible CFT duals for the entanglement wedge cross section have been suggested including the entanglement of purification \cite{Umemoto_2018,Nguyen2018}, odd entanglement \cite{Tamaoka_2019}, logarithmic negativity \cite{Kudler-Flam:2018qjo}, and reflected entropy \cite{Dutta:aa}. Of these the reflected entropy has the distinct advantage that it is more tractable to calculate in the boundary CFT. Holographically, the reflected entropy is the area of $b_{p\mathcal{P}}$ for the bipartite purification. In  \cite{Bao:2019zqc,Chu:2019etd} multipartite generalizations were considered where the multipartite reflected entropy was defined as the entanglement entropy of a particular purification. These choices involved a particular bipartition of several boundaries. For our construction since $\Sigma_{\mathcal{P}}$ is not homologous to the boundary of $\mathcal{P}$ we do not have this direct interpretation. Instead we are saying that topological data is being encoded in multipartite correlation of the boundary state. As of yet, there is no clear correspondence holographically between entanglement and such internal topology.
\begin{figure}[H]
\centering
\begin{tabular}{c}
\includegraphics[width=.3\textwidth]{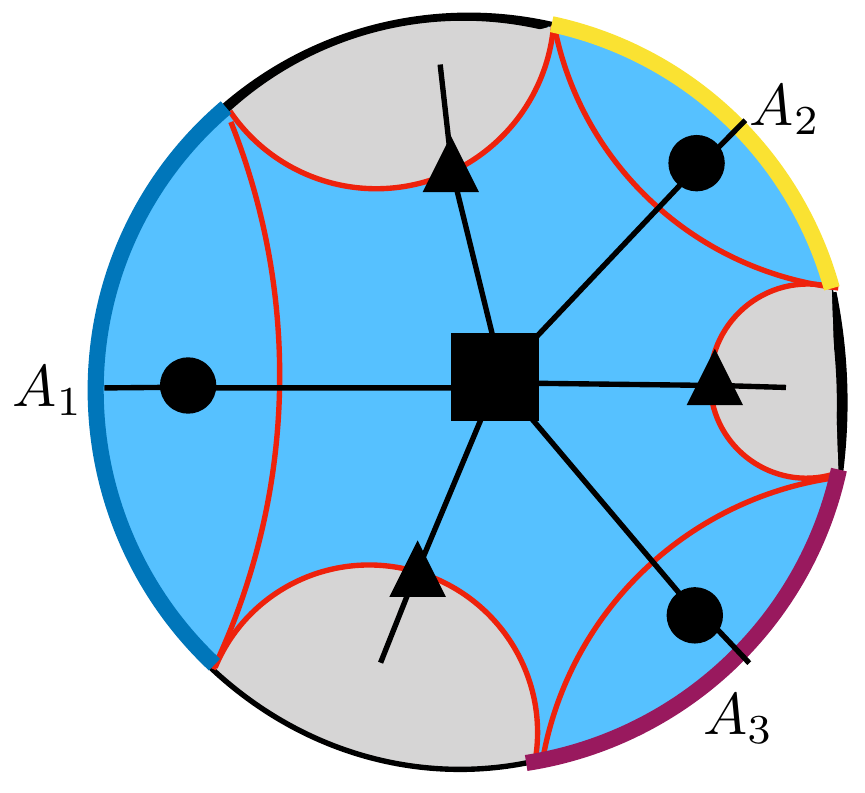}  \\
\includegraphics[width=.5\textwidth]{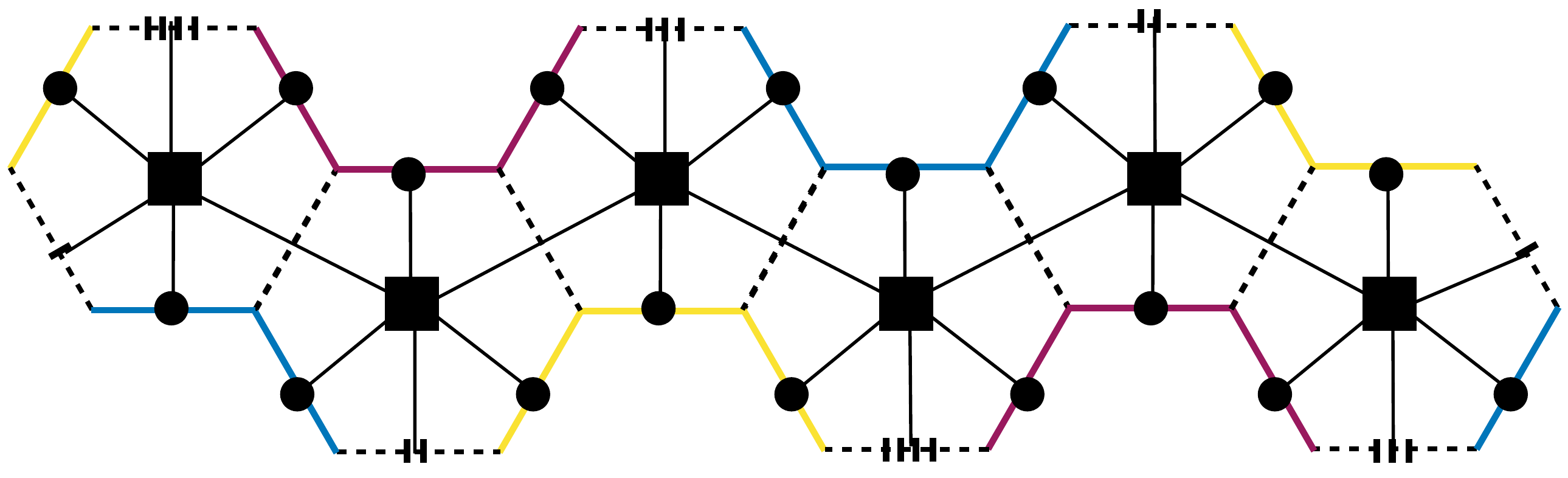}
\end{tabular}
\caption{\label{fig:tensornetwork}Top: Starting with a tensor for the internal region (square) we can push these legs out to the bounding RT surfaces. For those legs next to one of the boundary regions we can further push the leg out to the boundary (circle). We leave the remaining legs as uncontracted indices (triangle). Below: Using these tensor networks we create a new one by contracting legs where ever a gluing is performed. This gives us a new tensor network which approximates the boundary CFT state dual to $\mathcal{P}$.}
\end{figure}
\paragraph{Tensor Networks}
Using the construction of \cite{Bao:2018pvs} we can form a tree network of the entanglement wedge. This is done by choosing a collection of non-overlapping RT surfaces and looking at their dual graph. Working with our example of three boundary regions in a time slice of AdS$_{3}$ we have a tensor with six external legs: three, one for each boundary region, and three which form the components of $\mathcal{O}$. Taking our six copies of the tensor network we can glue the tensors together, as we did for the full geometry, by contracting tensors wherever we would perform a gluing of RT surfaces. From this we obtain a new tensor network which approximates the geometry of $\mathcal{P}$ (see figure \ref{fig:tensornetwork}). The remaining legs of the tensor network are now all associated with the new boundary regions of $\mathcal{P}$. Pushing a state to the boundary of the tensor network gives us a new CFT boundary state which is dual to our purified geometry. A similar procedure for the bipartite case was considered in \cite{Akers:2019gcv}.

\paragraph{Multiboundary wormholes}
Multiboundary wormholes are an important class of geometries for the study of holographic multipartite entanglement \cite{Krasnov:2000zq,Krasnov:2003ye,Skenderis:2009ju,Balasubramanian:2014hda}. Starting from the hyperbolic disc, Riemann surfaces of genus $g$ with boundaries $b$ are constructed by the identification of geodesics. For a given choice of $g$ and $b$ this construction gives rise to a spectrum of geometries which are labeled by the length of certain minimal surfaces: the wormhole throats homologous to each boundary and the internal moduli describing the topology behind the throats.

In our example above the interesting features of $\mathcal{P}$ were given by exactly these same quantities. This suggests a correspondence: given such a purification $\mathcal{P}$ we can construct a multiboundary wormhole geometry with the same spectrum of moduli. A connection between the multipartite entanglement wedge cross section and such geometries was already stated in \cite{Bao:2018ac,Bao:2018ab,Bhattacharya:2020ymw}. It is important to note that because we demand a fully connected entanglement wedge we are inherently looking at states which contain bipartite entanglement between any two pairs of boundary regions as diagnosed by the mutual information. This means we are only consider a portion of the full moduli space of wormhole geometries. Specifically this excludes any disconnected phases.

It would be very interesting to try and directly relate the internal moduli of multiboundary wormhole geometries to calculations of multipartite entanglement in the boundary dual.

\paragraph{Internal topology and measures of multipartite entanglement}
Given a collection of CFT boundaries in a particular entangled state it is well understood that the entanglement entropy holographically controls the values of the wormhole throats. This is simply a restatement of the RT formula. What is less understood is what controls the internal topology. Our construction of the purification for the entanglement wedge cross section suggests a partial answer by demonstrating a connection between measures of multipartite entanglement and the values of internal moduli. This is most clearly evidenced by the structure of maximal flows on $\mathcal{P}$ and suggests \emph{measures of multipartite entanglement are (at least in part) necessary to encode the internal topology of holographic spacetimes.}

It is well understood that bipartite entanglement and the connectedness of the entanglement wedge are diagnosed by entanglement entropy. Using the RT formula the area of the wormhole throats homologous to each boundary region are a direct measure
\begin{equation}\label{eq:RTbi}
S(A_{i})=\min_{m\in A_{i}}\area(m).
\end{equation}
As evidenced by our construction we believe the internal topology should be controlled in part by through measures of multipartite entanglement such as the multipartite entanglement wedge cross section
\begin{equation}\label{eq:RTmulti}
\Sigma=\min_{m\in \sigma}\area(m).
\end{equation}
It is our hope that this statement can be made more precise, though we leave such exploration to future work. We view this conjecture as complementary to claims made by \cite{Balasubramanian:2014hda,Bao:2015bfa,Akers:2019gcv} and others that multipartite entanglement plays a crucial role in holography.

\acknowledgments
The work of J.H. is supported in part by  the Simons Foundation through \emph{It from Qubit: Simons Collaboration on Quantum Fields, Gravity, and Information}. J.H would like to thank Charles Stine, Ning Bao, Matthew Headrick, Koji Umemoto, Newton Cheng, and Souvik Dutta for useful discussion. J.H would like to especially thank Matthew Headrick, Ning Bao, and Koji Umemoto for comments on an earlier version of this paper. J.H. would like to thank the Yukawa Institute for Theoretical Physics (YITP) and UC Davis for hospitality during various stages of this work.

\bibliographystyle{JHEP}
\bibliography{BT_MS}
\end{document}